\documentclass[twocolumn]{aastex631}

\usepackage{amsmath,amssymb}
\usepackage{xspace}
\usepackage{CJK}

\usepackage{hyperref}
\hypersetup{
    colorlinks=true,
    linkcolor=blue,
    filecolor=magenta,      
    urlcolor=cyan,
}

\usepackage{color}

\newcommand{\um}{\ensuremath{\mu\rm{m}}\xspace}
\newcommand{\kms}{\ensuremath{\rm{km\,s}^{-1}}\xspace}

\newcommand{\lfir}{\ensuremath{L_{\rm{FIR}}}\xspace}
\newcommand{\lir}{\ensuremath{L_{\rm{IR}}}\xspace}

\newcommand{\fagn}{\ensuremath{f_\mathrm{AGN}}\xspace}
\newcommand{\soh}{\ensuremath{S_\mathrm{120\um}}\xspace}
\newcommand{\scii}{\ensuremath{S_\mathrm{160\um}}\xspace}
\newcommand{\Tdust}{\ensuremath{T_{\rm{dust}}}\xspace}

\newcommand{\rdust}{\ensuremath{r_{\rm{dust}}}\xspace}

\newcommand{\MBH}{\ensuremath{M_{\rm{BH}}}\xspace}
\newcommand{\Msol}{\ensuremath{\rm{M}_\odot}\xspace}
\newcommand{\Lsol}{\ensuremath{\rm{L}_\odot}\xspace}
\newcommand{\LAGN}{\ensuremath{L_{\rm{AGN}}}\xspace}
\newcommand{\LSF}{\ensuremath{L_{\rm{SF}}}\xspace}

\newcommand{\vmax}{\ensuremath{v_\mathrm{max}}\xspace}
\newcommand{\vfifty}{\ensuremath{v_{50}}\xspace}
\newcommand{\vef}{\ensuremath{v_{84}}\xspace}
\newcommand{\cii}{[C{\scriptsize II}]\xspace}

\newcommand{\oiii}{[O{\scriptsize III}]\xspace}

\newcommand{\arc}{\ensuremath{''}\xspace}

\newcommand{\Mout}{\ensuremath{M_{\mathrm{out}}}\xspace}
\newcommand{\Rout}{\ensuremath{R_{\mathrm{out}}}\xspace}
\newcommand{\vout}{\ensuremath{v_{\mathrm{out}}}\xspace}
\newcommand{\tout}{\ensuremath{t_{\mathrm{out}}}\xspace}

\newcommand{\etaout}{\ensuremath{\eta_{\mathrm{out}}}\xspace}
\newcommand{\Mdot}{\ensuremath{\dot{M}_{\mathrm{out}}}\xspace}
\newcommand{\pdot}{\ensuremath{\dot{p}_{\mathrm{out}}}\xspace}
\newcommand{\Edot}{\ensuremath{\dot{E}_{\mathrm{out}}}\xspace}

\newcommand{\EWvth}{\ensuremath{\mathrm{EW}_{v < -200}}\xspace}

\shortauthors{J.~S.~Spilker, et~al.}
\shorttitle{Cold Quasar Outflows}

\begin{document}
\begin{CJK*}{UTF8}{gbsn}

\defcitealias{spilker18a}{S18}
\defcitealias{spilker20a}{S20a}
\defcitealias{spilker20b}{S20b}
\defcitealias{herreracamus20}{HC20}

\title{Direct Evidence for AGN Feedback from Fast Molecular Outflows in Reionization-Era Quasars}

\correspondingauthor{Justin S. Spilker}
\email{jspilker@tamu.edu}

\author[0000-0003-3256-5615]{Justin~S.~Spilker}
\affiliation{Department of Physics and Astronomy and George P. and Cynthia Woods Mitchell Institute for Fundamental Physics and Astronomy, Texas A\&M University, 4242 TAMU, College Station, TX 77843-4242, US}

\author[0000-0002-6184-9097]{Jaclyn~B.~Champagne}
\affiliation{Steward Observatory, University of Arizona, 933 N. Cherry Ave., Tucson, AZ 85721, US}

\author[0000-0003-3310-0131]{Xiaohui~Fan}
\affiliation{Steward Observatory, University of Arizona, 933 N. Cherry Ave., Tucson, AZ 85721, US}

\author[0000-0001-7201-5066]{Seiji~Fujimoto}
\affiliation{Department of Astronomy, University of Texas at Austin, Austin, TX 78712, US}

\author[0000-0001-5434-5942]{Paul~P.~van~der~Werf}
\affiliation{Leiden Observatory, Leiden University, PO Box 9513, 2300 RA Leiden, Netherlands}

\author[0000-0001-5287-4242]{Jinyi~Yang}
\affiliation{Steward Observatory, University of Arizona, 933 N. Cherry Ave., Tucson, AZ 85721, US}

\author[0000-0002-5367-8021]{Minghao~Yue}
\affiliation{MIT Kavli Institute for Astrophysics and Space Research, 77 Massachusetts Ave., Cambridge, MA 02139, US}

\begin{abstract}

Galactic outflows driven by rapidly-accreting quasars at high redshift are widely expected to play a key role in the short- and long-term future evolution of their host galaxies. Using new and archival ALMA data, we observed the OH 119\,\um doublet lines in order to search for cold molecular outflows in a sample of 11 unobscured, IR-luminous quasars at $z>6$. This represents the first survey for molecular winds in reionization-era quasars, and we detect unambiguous outflows in 8/11 (73\%) of the quasars. The outflows we find are substantially faster, by $\approx$300\,\kms on average, than outflows observed in a roughly co-eval sample of non-quasar IR-luminous galaxies, suggesting that the AGN drive the winds to higher velocities. On the other hand, the implied molecular outflow rates are relatively modest given the high luminosities, suggesting typical mass loading factors $\sim$0.5 in the cold gas. The outflows are consistent with expectations for momentum-driven winds regardless of the driving source, but the kinetic energy in the outflows suggests that the AGN must be at least partially responsible for driving the winds. Accordingly, we find trends between the outflow properties and the Eddington ratio of the black hole accretion, though this may be linked to the underlying trend with AGN luminosity. We find that the kinetic power carried in the cold outflow phase is typically only $\sim0.1$\% of the total AGN luminosity. Our study provides evidence in favor of AGN feedback on the cold molecular gas in $z>6$ quasar host galaxies, demonstrating that cold outflows are very common and powerful in the most extreme reionization-era quasars. 

\end{abstract}

\section{Introduction} \label{intro}

Over the past few decades, it has become clear that any successful model of galaxy evolution requires one or more forms of feedback that enable the self-regulation of galaxy growth. While `feedback' itself refers to a diverse set of physical processes, collectively these processes influence the shape and normalization of almost all galaxy scaling relations, linking together star formation, stellar mass growth, supermassive black hole (SMBH) accretion, galaxy metallicity, the circumgalactic medium, and more \citep[for recent reviews, see e.g.][]{somerville15,veilleux20,harrison24}. While the energy and momentum injected into the galaxy interstellar medium (ISM) by star formation is relevant across all mass scales, there is broad theoretical consensus that the most massive systems require additional feedback energy from quasar-mode SMBH accretion \citep[e.g.][]{hopkins08,vogelsberger14,morganti17}.

Among the most obvious direct evidence for feedback is the prevalence of large-scale outflows or winds launched from galaxies at all mass scales. The gas in outflows spans many orders of magnitude in temperature and density from diffuse ionized plasma to cold, dense molecular gas \citep[e.g.][]{veilleux20}. For both better and worse, this makes different phases of outflows detectable across all wavelengths, and evidence of outflows can be subtle. Historically, the warm ionized and neutral atomic phases have been the most studied because they have observable signatures at rest-frame or redshifted optical wavelengths \citep[e.g.][]{heckman90,rupke05,weiner09}. From multiwavelength accounting, however, the total outflowing mass seems to be dominated by cold, dense, molecular gas \citep[e.g.][]{rupke13,fiore17,fluetsch19}. Because stars also form from this cold gas, molecular outflows are of particular interest.

In the early universe, limitations in sensitivity and/or resolution make detecting outflows in typical galaxies very difficult even with powerful telescopes such as ALMA and JWST \citep[e.g.][]{herreracamus21,belli24}. Consequently, most high-redshift outflow studies have focused on galaxy populations that should show the effects of feedback most strikingly, including massive star-forming galaxies and quasars. Both populations are also often heavily dust-obscured and therefore bright at far-infrared wavelengths. Out to $z \sim 2$, the ionized/neutral phase outflow tracers in the rest-frame optical are still accessible from the ground, so many surveys have targeted these tracers, especially in quasar host galaxies \citep[e.g.][]{perna15,zakamska16,circosta18,scholtz20,vayner21b}. These same outflow tracers are now available at even higher redshifts with JWST, and early results have begun to characterize the warm phase of outflows in diverse populations of galaxies, both with and without active galactic nuclei (AGN) \citep[e.g.][]{carniani24,park24,vayner24,wang24}.

The coldest phase of outflows, typically traced by lines of the CO, OH, or H$_2$O molecules, or the \cii 158\,\um atomic fine structure line, remains challenging to detect at high redshifts. While lower-redshift studies often have the advantage of superior physical resolution to help distinguish outflowing gas from the rest of the galaxy \citep[e.g.][]{lamperti22}, high-redshift studies typically only detect outflows through high-velocity line wings in emission (CO, \cii) or absorption (OH, H$_2$O). Along with the overall faintness of the high-velocity wings, the result is that many cold outflow studies that rely on emission lines at high redshift are controversial. Stacking analyses of the \cii line in samples of up to a few dozen $z>6$ quasars, for example, have come to opposite conclusions about whether any broad high-velocity component is present, even though many of the same quasars are in common \citep{decarli18,bischetti19,stanley19,novak20}. The same controversy arises even in individual objects: there are a growing number of high-redshift quasars and non-quasars with unambiguous outflows detected via OH absorption lines that show few signs of broad \cii line wings \citep{spilker20a,salak24}, and at least some claims of high-velocity \cii wings \citep{maiolino12,cicone15} have turned out to be spurious \citep{meyer22}.

Despite these challenges, the properties of cold outflows from quasars are beginning to come into view out to $z\sim2$ \citep[e.g.][]{nesvadba11,stacey22,butler23b}, including some quasars with multiphase outflow constraints from multiwavelength data \citep[e.g.][]{brusa18,vayner21a}. While there is still a great deal to learn about quasar molecular outflows at these redshifts, the general consensus is that the AGN do play a role in driving the cold and dense outflow phase, removing cold gas at a rate comparable to the star formation rate (SFR).

Beyond $z>4$, CO line wings are typically too faint to detect with ALMA, and \cii seems to be an unreliable outflow tracer, at best, for reasons that remain unclear. Fortunately, at these redshifts the strongest OH lines redshift to frequencies accessible with ALMA. The ground-state OH $^2 \Pi_{3/2}$ $J = 3/2 \rightarrow 5/2$ doublet transitions at rest-frame $\approx$119\,\um were extensively observed by the Herschel Space Observatory in the nearby universe \citep[e.g.][]{sturm11,spoon13,veilleux13,stone16,gonzalezalfonso17}, providing a low-redshift comparison sample with like observables. The first modest-size sample of high-redshift OH 119\,\um observations targeted IR-luminous dusty star-forming galaxies (DSFGs) with no evidence for unobscured or obscured AGN (\citealt{spilker18a,spilker20a,spilker20b}, hereafter \citetalias{spilker20a,spilker20b}). These studies showed that molecular outflows are very common in high-redshift DSFGs, have outflow rates near but somewhat below the SFRs, and have energetics consistent with being launched and powered by star formation.

The growing population of $z>6$ unobscured quasars selected from wide-area surveys are also attractive targets for high-redshift feedback studies \citep[see][for a recent review of these sources]{fan23}. By this time, several dozen of these quasars have been detected in the far-IR \cii line and underlying dust continuum by ALMA. The implied IR luminosities mean that, even though the AGN has cleared the line of sight to the nucleus, the host galaxies contain vast quantities of dust heated by star formation or possibly the AGN itself \citep[e.g.][]{venemans18}.  For the $z>6$ quasars with the brightest far-IR continuum, OH 119\,\um absorption spectroscopy is eminently feasible, and OH has been observed in a handful of them so far (\citealt{herreracamus20}, hereafter \citetalias{herreracamus20}; \citealt{butler23,salak24}). These studies collectively detected tentative or clear outflows in 4/5 targets, but the sample is still too small to make robust comparisons to either low-redshift quasars or high-redshift non-quasar DSFGs. 

Here we begin to move beyond these small-sample studies in an attempt to understand the properties of molecular outflows in $z>6$ quasars on a statistical basis. We present new ALMA observations of OH 119\,\um in 8 quasars chosen to have bright far-IR continuum emission so as to make a modest sample size achievable in a short observing program. We reanalyze the previous literature observations to make self-consistent comparisons, focusing mainly on comparisons to the $z\sim5$ DSFG sample of \citetalias{spilker20a} and low-redshift IR-luminous galaxies. Section~\ref{data} describes our sample selection and observations. Section~\ref{analysis} outlines our analysis methods, including how we estimate outflow physical properties (outflow rates, etc.) from the observations. Our main results are presented in Section~\ref{results}, and we conclude in Section~\ref{conclusions}. We assume a flat $\Lambda$CDM cosmology with $\Omega_m=0.307$ and $H_0=67.7$\,\kms\,Mpc$^{-1}$ \citep{planck15}.

\section{Sample and Observations} \label{data}

\subsection{Quasar Sample Selection} \label{selection}

We selected a sample of $z > 6$ quasars that had pre-existing detections of the \cii 158\,\um line and underlying $\approx$160\,\um continuum from ALMA. The parent sample comprises a few dozen quasars from the literature, both modest-size surveys and single- or few-object studies \citep{gallerani12,wang16,venemans18,bischetti19,hashimoto19,andika20,venemans20,yang21,yue21}. Our primary selection criteria were that: (i) the quasar redshift places the OH 119\,\um doublet at frequencies of favorable atmospheric transmission, and (ii) that the 120\,\um continuum was predicted to be bright enough to be able to detect typical 10\% OH absorption depths in $\lesssim$3\,h of total ALMA observing time. In practice, we required $\scii > 4$\,mJy, with an expectation that $\soh / \scii \gtrsim 1.5$ for typical cold dust temperatures $\Tdust \gtrsim 35$\,K. This restriction limits our sample to among the most IR-luminous quasars known at $z>6$, with $\log \lfir/\Lsol \gtrsim 12.5$. Our results may not be applicable to less-luminous sources. Our goal was not to assemble a representative sample of $z>6$ quasars, but rather to assemble a sample large enough to provide the first constraints on the outflow detection rate among such objects, in order to motivate future observations of fainter, more typical sources.

The sample we selected for ALMA observations comprises 8 quasars drawn from the literature. Three additional quasars with existing OH observations also met our selection criteria, and we include these sources in our analysis: J1319+0950 \citepalias{herreracamus20}, P036+03 \citep{butler23}, and J2054-0005 \citep{salak24}. \citet{herreracamus20} claim a tentative outflow in J1319+0950. For the purposes of our work we do consider this object to have a confirmed outflow, 
but our conclusions are not significantly affected by the inclusion or exclusion of this source. Our sample also includes one gravitationally-lensed quasar, J0439+1634; for this source we correct for the lensing magnification $\mu = 4.5$ estimated from modeling of ALMA 160\,\um continuum imaging \citep{yue21}.

\subsection{ALMA OH Observations} \label{almaobs}

ALMA observations were carried out in program 2021.1.00443.S (PI: Spilker) from April -- September 2022, and are summarized in Table~\ref{tab:almaobs}. The OH 119\,\um doublet is redshifted to ALMA Band~7, $\nu_{\rm{obs}} \approx 330-358$\,GHz.  The OH doublet was covered using two slightly overlapping 1.875\,GHz basebands with 7.8\,MHz channels, providing $\approx$3100--3400\,\kms of total spectral coverage. We placed the basebands to cover up to 1600\,\kms blueward of the short-wavelength OH doublet component. The doublet features are separated by $\approx$520\,\kms, leaving 1000--1300\,\kms to help constrain the continuum level redward of the lower-frequency doublet component. We placed two additional basebands in the opposite sideband to constrain the continuum, choosing either the lower or upper sideband based on the atmospheric transmission given the IF range of the ALMA Band~7 receivers.  On-source integration times ranged from 50--100\,min, set by a combination of the predicted continuum brightness and the fraction of the requested time actually executed by ALMA. 

We used compact array configurations to avoid over-resolving the target quasars. With typical baseline ranges of $\approx$15--500\,m, the data reach spatial resolution $\approx$0.6'' when imaged using natural weighting of the visibilities. The maximum angular scale recoverable by the data is $>$5'', much larger than the expected sizes of high-redshift quasars. For all quasars, we performed a single round of phase-only self-calibration, with the solution interval equal to the scan length. The self-calibration improved the peak continuum S/N by an average (median) factor of $\approx$70\% ($\approx$40\%). Continuum images were generated from the alternate-sideband (continuum-only) data, all of which reach very high S/N. 

None of the quasars are pointlike at the depth and resolution of our data, but all are at-best moderately resolved (including the lensed quasar J0439+1634). We use the CASA \texttt{imfit} task to fit elliptical Gaussian profiles to the continuum images of each quasar, and we measure the 120\,\um continuum flux density and source sizes from the best-fit profiles. The (circularized FWHM) source sizes, deconvolved from the synthesized beam, range from 0.2--0.5''. We detect additional continuum sources in some fields (J0305-3150, J0525-2406, P083+11, P231-20), at least some of which were previously known. For these fields, we include and simultaneously fit these additional sources together with the target quasars. Of the three fields with companions spectroscopically confirmed at the quasar redshift (J0305-3150 source `C3', \citealt{venemans19}; P083+11 southern companion, \citealt{andika22}; P231-20 companion, \citealt{decarli17}), we detect OH only in the P231-20 companion; it has an absorption profile tentatively blueshifted by $\approx$400\,\kms.

We created cubes of both the OH-containing and alternate sidebands at 75\,\kms spectral resolution, again using natural weighting. We extract source spectra by turning again to the \texttt{imfit} task, this time fixing the source position, size and shape to the best-fit values measured in the continuum images, leaving only the flux density in each channel as a free parameter. This is justified because OH 119\,\um is typically seen in absorption, so it cannot extend beyond the continuum emission. The gas densities required to produce OH in emission are so high that any OH emission likewise should not be more extended than the continuum arising from the larger host galaxy. Spectra of our sample quasars are shown in Figure~\ref{fig:ohspectra}, and Figure~\ref{fig:images} displays images of the continuum and OH emission and absorption, if present.

Two of our quasar sources, J2310+1855 and P183+05, were also previously observed in the OH 119\,\um transitions \citep{butler23}.\footnote{We did not re-observe the third source from \citet{butler23}, P036+03.} We re-observed these two sources, but shifted the spectral window placement to extend to higher frequencies than the previous data by about 0.8\,GHz ($\approx$800\,\kms). We noticed a large discrepancy in the flux level between the previous data and our own and so did not combine these data with our own. We measure continuum flux densities higher by 50\% (P183+05) and 250\% (J2310+1855) than \citet{butler23}. We downloaded the previous data from the ALMA archive and successfully reproduced the \citet{butler23} measurements to within the stated uncertainties; the discrepancy appears to be in the data and not in any analysis steps. The origin of the mismatched flux scale is not clear, but is unlikely to be caused by differences in weather conditions (assuming accurate flux calibration), array configuration (similar between datasets), or self-calibration (impacting peak but not necessarily integrated fluxes). Between the two discrepant measurements, we believe our new values to be correct for two reasons. First, the ALMA sensitivity calculator predictions agree with the noise levels we measure in our data but predicts much higher noise than measured in the older data. Second, the absolute flux scale of the \citet{butler23} data results in continuum flux densities significantly lower than expected based on other ALMA observations of these sources (see Appendix~\ref{appsedfits}). We note that most of the derived quantities in the \citet{butler23} analysis rely on the OH equivalent widths rather than absolute fluxes, and so are not necessarily impacted by this flux scale mismatch (though our data better constrains the outflow velocity thanks to the wider frequency coverage, which impacts estimates of the outflow rate, etc.).

\begin{deluxetable*}{lcccccc}
\tablecaption{Summary of New ALMA OH Observations \label{tab:almaobs}}
\tablehead{
\colhead{Source} &
\colhead{R.A.}   &
\colhead{Decl.}  &
\colhead{t$_{\mathrm{on\,source}}$ (min.)} &
\colhead{Beam (arcsec)} &
\colhead{$\sigma_{\mathrm{cont}}$ ($\mu$Jy/beam)} &
\colhead{$\sigma_{75}$ (mJy/beam)}}
\startdata
J0305-3150 & 03:05:16.92 & $-$31:50:55.8 & 94 & 0.53$\times$0.75 & 28 & 0.27 \\
J0439+1634 & 04:39:47.10 & $+$16:34:15.7 & 49 & 0.60$\times$0.70 & 27 & 0.20 \\
J0525-2406 & 05:25:59.67 & $-$24:06:23.0 & 50 & 0.53$\times$0.68 & 19 & 0.16 \\
J2310+1855 & 23:10:38.90 & $+$18:55:19.8 & 84 & 0.64$\times$0.89 & 19 & 0.17 \\
P009-10    & 00:38:56.53 & $-$10:25:54.1 & 100& 0.56$\times$0.70 & 18 & 0.17 \\
P083+11    & 05:35:20.89 & $+$11:50:53.8 & 100& 0.59$\times$0.64 & 17 & 0.13 \\
P183+05    & 12:12:26.97 & $+$05:05:33.5 & 99 & 0.62$\times$0.65 & 19 & 0.16 \\
P231-20    & 15:26:37.84 & $-$20:50:00.7 & 100& 0.56$\times$0.73 & 16 & 0.14 \\
\enddata
\tablecomments{Synthesized beam sizes and sensitivities are given for self-calibrated, naturally-weighted images. The line sensitivity $\sigma_{75}$ is given for the 75\,\kms channel closest to the upper OH rest frequency.}
\end{deluxetable*}

\subsection{Ancillary Data} \label{ancillary}

For all sources, we use the ALMA \cii observations to define the systemic redshift of each quasar. Because these data come from a variety of programs and teams, the spatial resolution, frequency coverage, and sensitivity vary widely. Some targets have been observed in \cii in multiple programs; in those cases we preferred the datasets with the lowest spatial resolution and widest velocity coverage on either side of the \cii line. We do not perform a detailed analysis of the \cii data here.

Some quasars in our study have extensive additional far-IR observations, while others have few or no additional measurements beyond the \cii and OH observations described previously. We searched the literature and the ALMA archive to assemble all the available far-IR photometry for our sample quasars. There are disappointingly few Herschel observations of these sources, which makes robust constraints on the total IR luminosity \lir (integrated from 8--1000\,\um rest-frame) difficult. Of the 11 quasars in our sample (plus the literature quasars), J2310+1855 and J2054-0005 are both detected in Herschel/PACS and SPIRE imaging \citep{leipski14,tripodi22}. P183+05 and J1319+0950 were observed by SPIRE but not detected (\citealt{pappalardo15}; R.~Wang, priv. comm.); for these sources we assume upper limits on the 250, 350, and 500\,\um flux density equal to the SPIRE confusion limits. The photometric measurements are tabulated in Appendix~\ref{appsedfits}, and our SED fitting methods described in Sec.~\ref{sedfitting}.

\section{Analysis} \label{analysis}

\subsection{Spectral Analysis and Outflow Identification} \label{specfitting}

\begin{figure*}[t]
\begin{centering}
\includegraphics[width=0.49\textwidth]{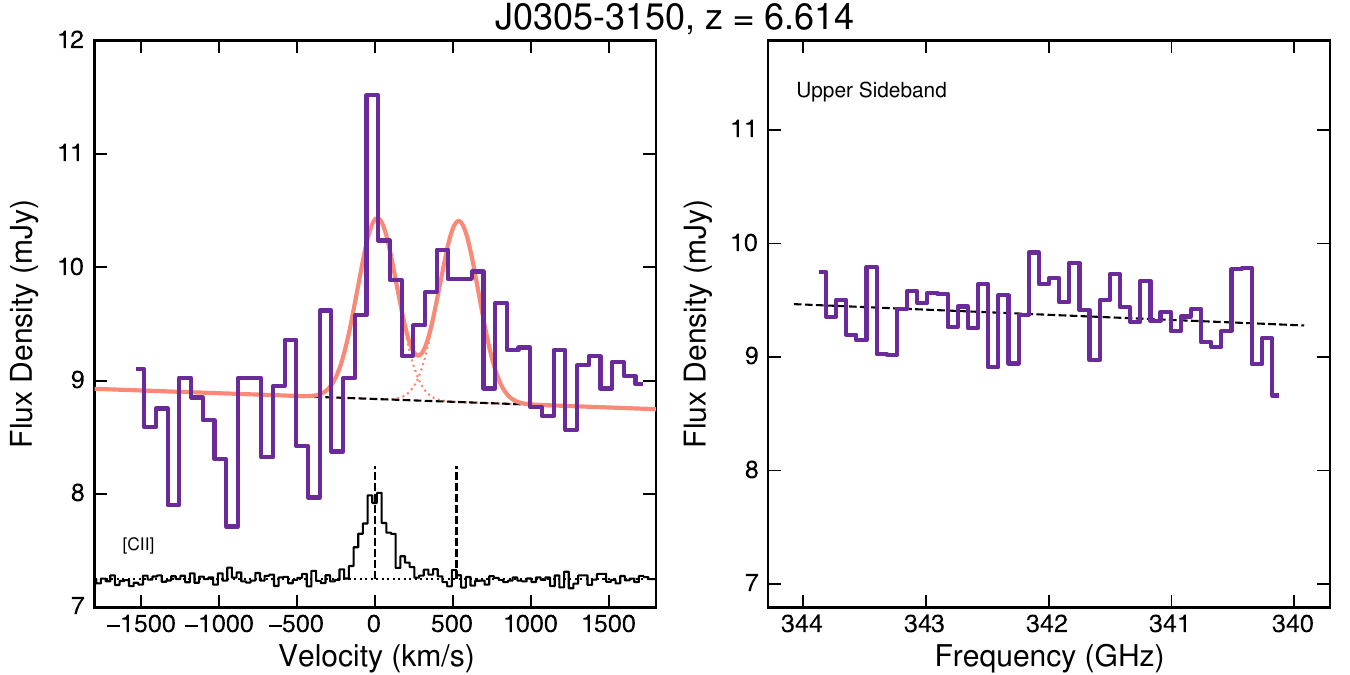}
\includegraphics[width=0.49\textwidth]{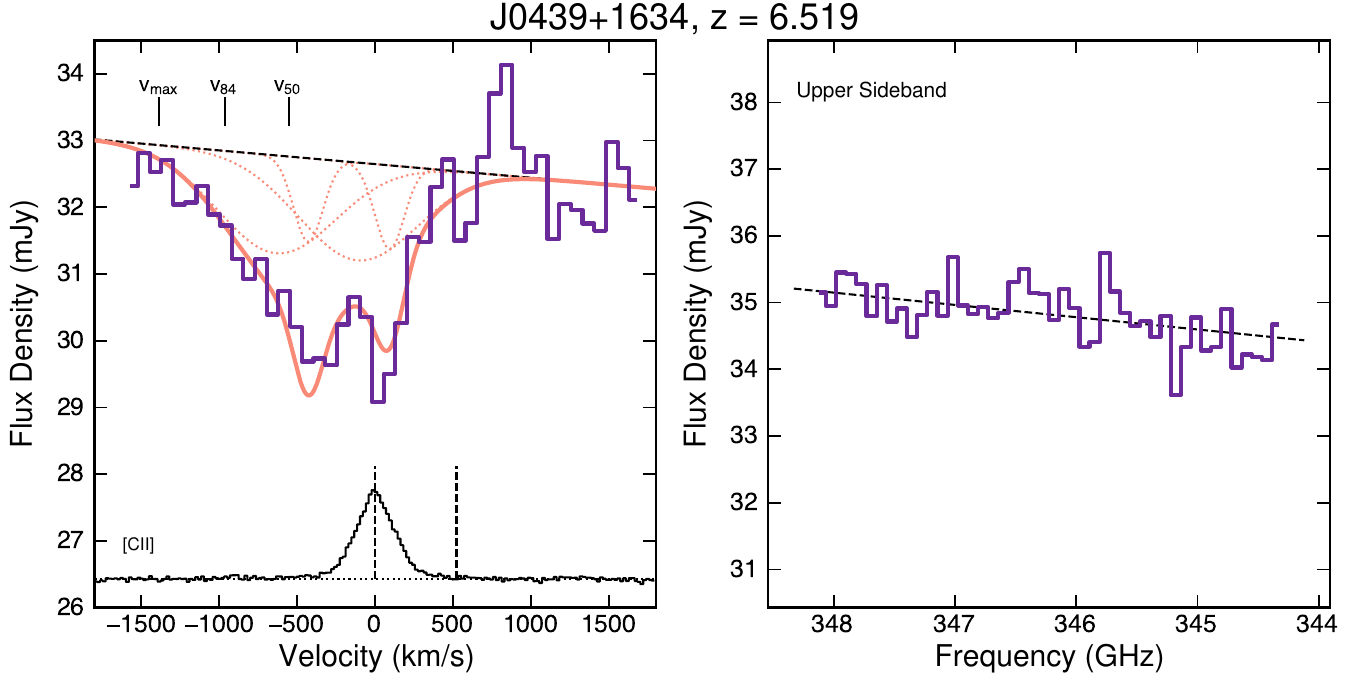}
\includegraphics[width=0.49\textwidth]{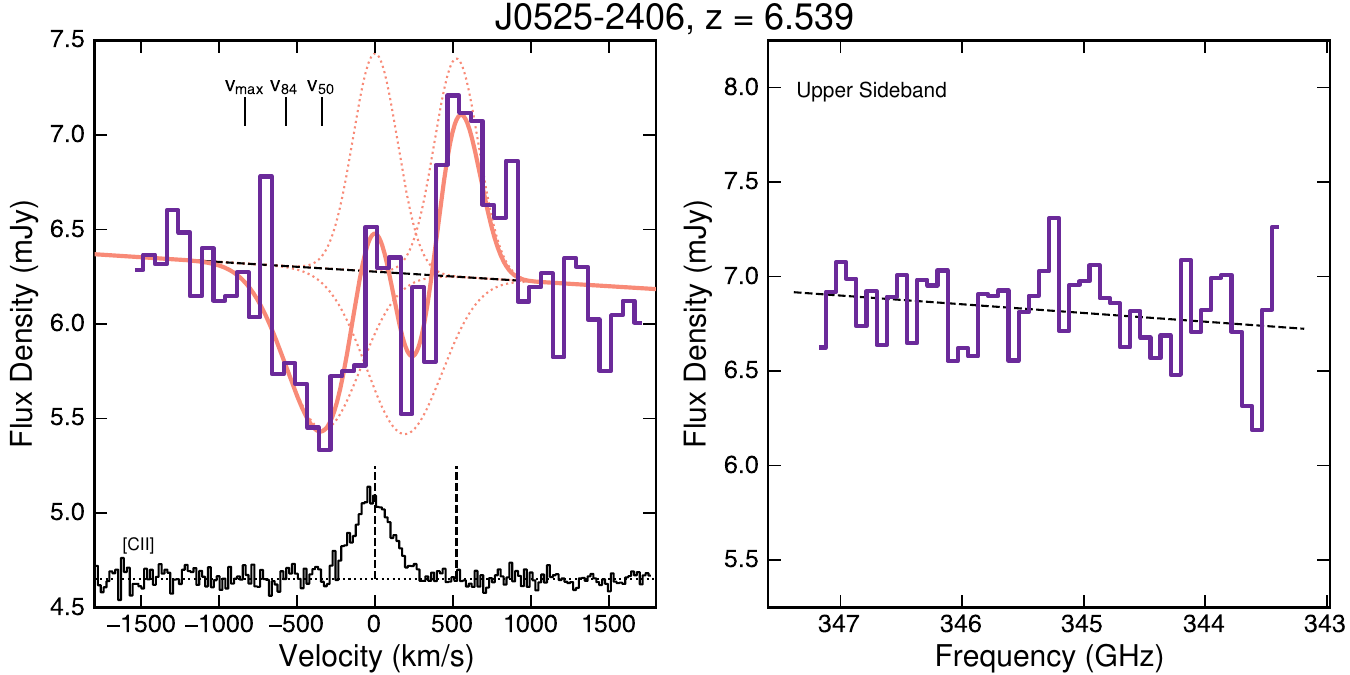}
\includegraphics[width=0.49\textwidth]{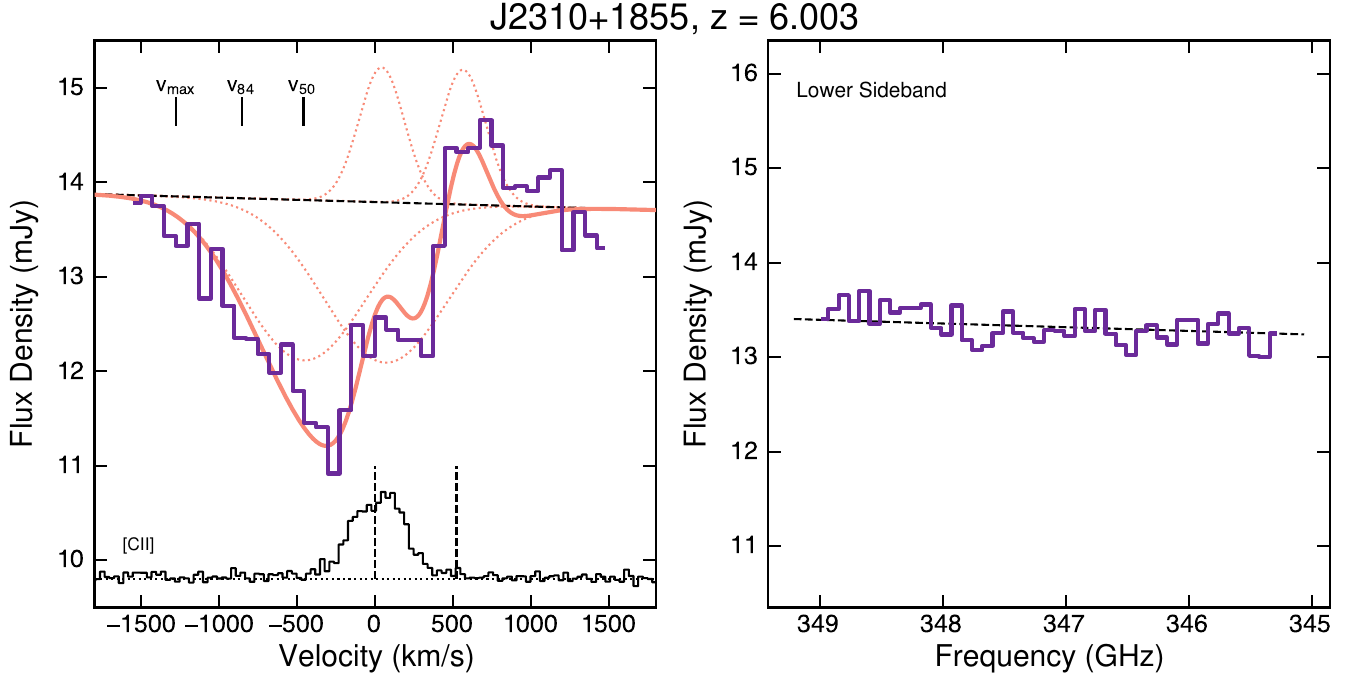}
\includegraphics[width=0.49\textwidth]{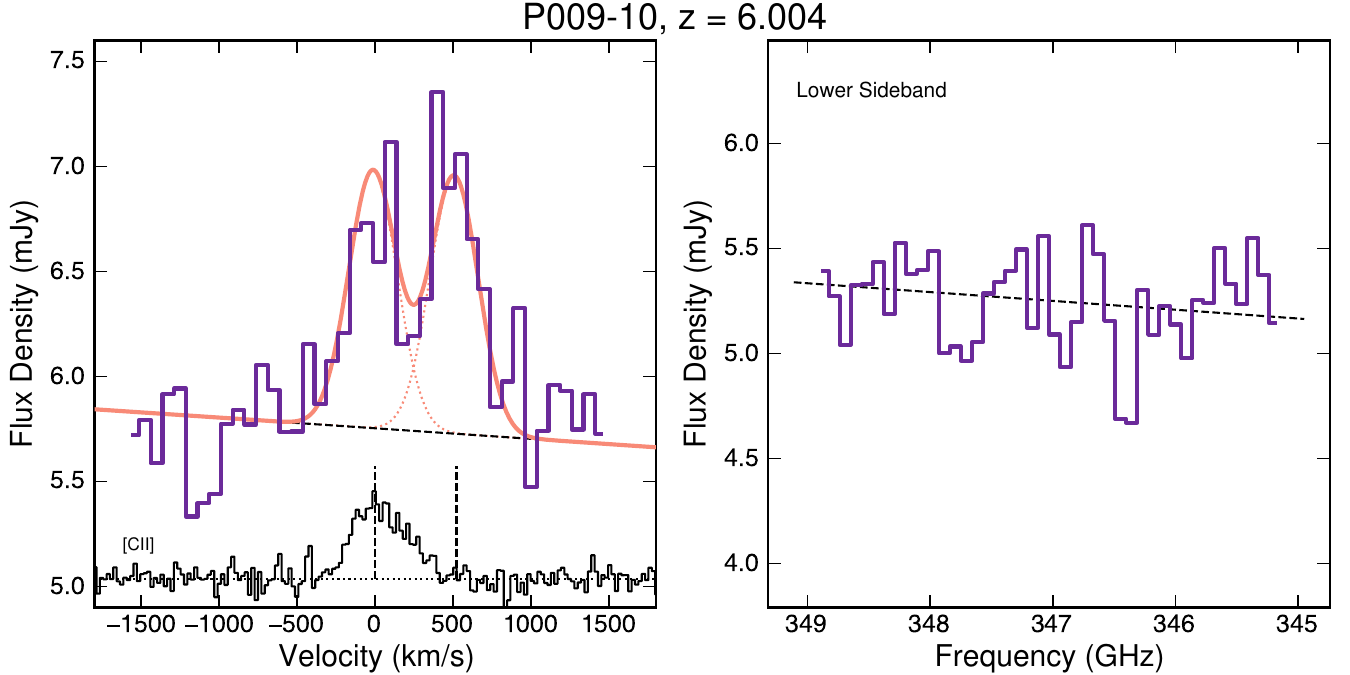}
\includegraphics[width=0.49\textwidth]{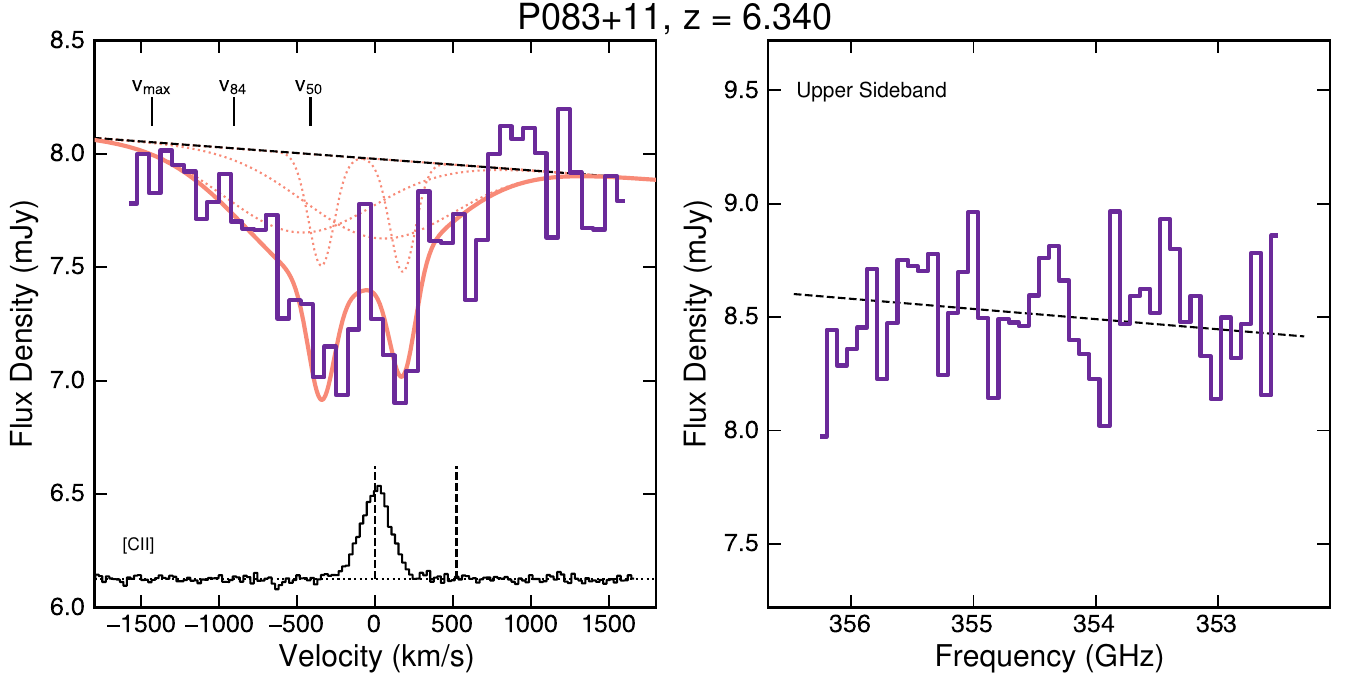}
\includegraphics[width=0.49\textwidth]{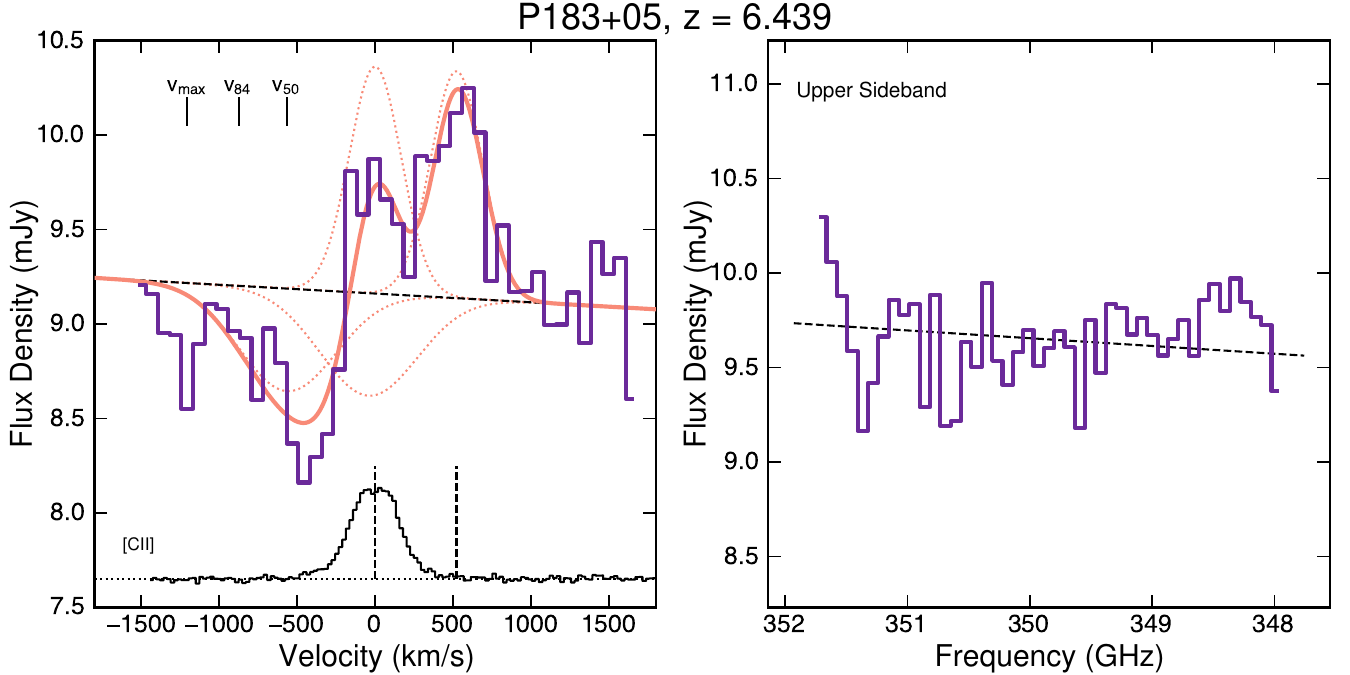}
\includegraphics[width=0.49\textwidth]{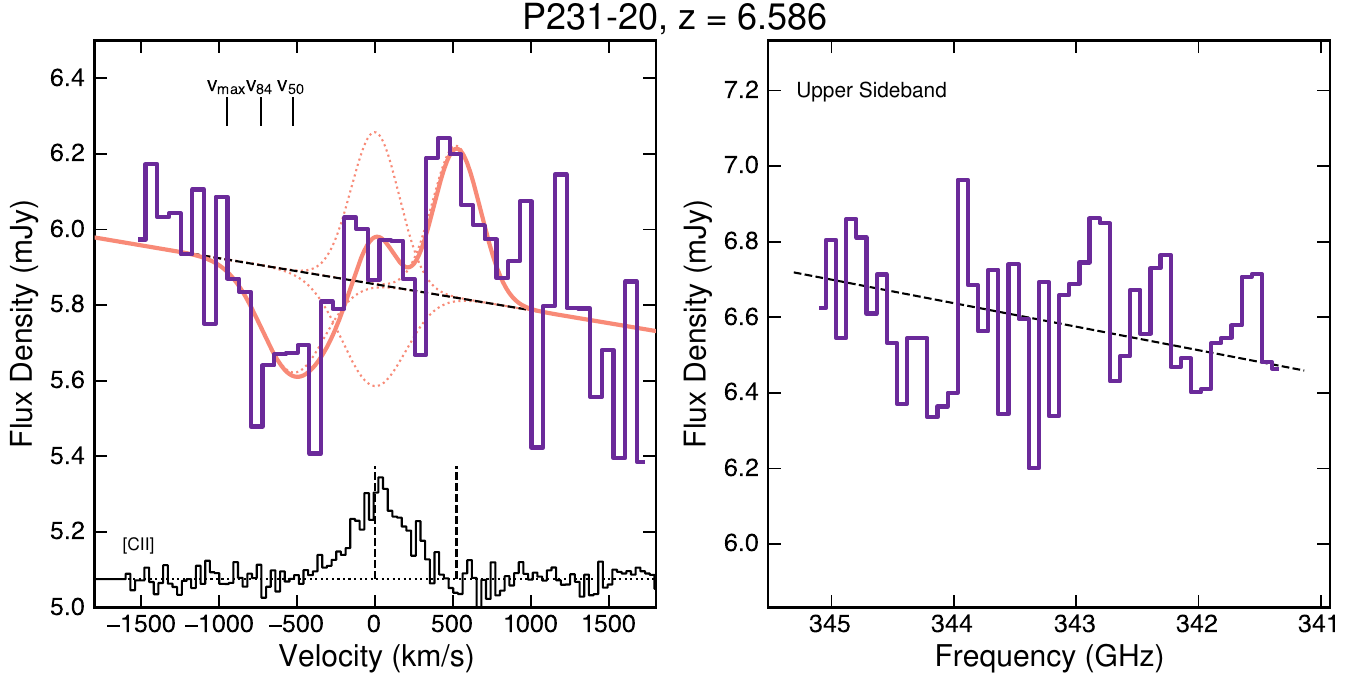}\\
\end{centering}
\caption{
OH 119\,\um spectra for our $z>6$ quasar sample (navy) and spectral fits (peach) using one or two pairs of Gaussian profiles. The total OH line fit is shown as a solid peach line, with individual line components shown as thin dotted profiles. The continuum level (black dashed lines) is constrained from a joint fit to both sidebands of the ALMA data. For sources with OH absorption, we mark the outflow velocity metrics \vfifty, \vef, and \vmax with vertical ticks. We show the \cii spectra of each source (with arbitrary offset and normalization) to indicate the systemic emission. The doublet OH components are separated by 520\,\kms, marked by vertical dashed lines.
}\label{fig:ohspectra}
\end{figure*}

The OH spectra of our quasar sample objects are shown in Figure~\ref{fig:ohspectra}. We see a variety of OH profiles in our sources, including pure absorption, pure emission, and combined emission/absorption. This is qualitatively similar to past work at low redshift \citep[e.g.][]{veilleux13}, but contrasts with the $z\sim5$ non-quasar dusty galaxies of \citetalias{spilker20a}, in which only pure absorption was observed.

To interpret the data, we perform a simple least-squares fitting of either one or two pairs of Gaussians to the OH spectra, depending on the complexity of the spectra. We fix the amplitude of each Gaussian in a pair to be the same, since these OH transitions are $\Lambda$ doublets, and we fix the velocity separation within a pair to 520\,\kms based on the known frequencies of the transitions. We ignore the additional hyperfine splitting of each doublet component because this $\approx$5\,\kms splitting is far below the spectral resolution of the data. Because the OH profiles take up a large fraction of the total bandpass, we also use the alternate line-free sideband to help constrain the continuum level. We assume a simple linear continuum with frequency, and jointly fit both the OH-containing and alternate sideband. The slope of the continuum is not particularly well constrained by a single sideband, but the gap in frequency between sidebands results in noticeably different continuum levels between them. For sources that exhibit both emission and absorption, we follow \citet{veilleux13} and constrained the amplitudes to avoid having arbitrarily large emission components be canceled by equal-and-opposite absorption components. We stress that our fits are certainly not unique, but they do capture the features of the observed OH spectra.

From the spectral fits, we follow standard practice and derive the OH equivalent width. We also derive `non-parametric' measures of the OH velocities, namely \vfifty, \vef, and \vmax, taken to be the velocities above which 50\%, 84\%, and 98\% of the OH absorption takes place. We experimented with alternative spectral fitting procedures (varying the number of Gaussian pairs, requiring or preventing an emission component, etc.) and found that these derived properties are robust to changes in the methodology. We also measure the OH equivalent widths at velocities more blueshifted than 200\,\kms, which we use to estimate outflow rates \citepalias[see also][]{spilker20b}. These and other useful sample properties are given in Table~\ref{tab:ohfits}.


\begin{deluxetable*}{lccccccccccc} 
\tablecaption{Quasar sample properties and quantities derived from the OH spectra \label{tab:ohfits}}
\tablehead{
\colhead{Source} & 
\colhead{$z_{\cii}$} & 
\colhead{\lfir} & 
\colhead{\LSF} & 
\colhead{\LAGN} & 
\colhead{$S_{\mathrm{119\um}}$} & 
\colhead{\rdust} & 
\colhead{Outflow?} & 
\colhead{\vfifty} & 
\colhead{\vef} & 
\colhead{\vmax} & 
\colhead{\EWvth} \\ 
\colhead{} & 
\colhead{} & 
\colhead{10$^{12}$\,\Lsol} & 
\colhead{10$^{12}$\,\Lsol} & 
\colhead{10$^{12}$\,\Lsol} & 
\colhead{mJy} & 
\colhead{kpc} & 
\colhead{} & 
\colhead{\kms} & 
\colhead{\kms} & 
\colhead{\kms} & 
\colhead{\kms} 
} 
\startdata 
J0305-3150 & 6.614 & 7.0$\,^{+1.9}_{-1.1}$ & 7.5$\,^{+4.5}_{-2.5}$ & 25$\,^{+10}_{-7}$ & 8.83 $\pm$ 0.07 & 1.32 & N & --- & --- & --- & --- \\ 
J0439+1634\tablenotemark{a} & 6.519 & 5.0$\,^{+1.1}_{-0.7}$ & 5.4$\,^{+2.6}_{-1.6}$ & 11$\,^{+5}_{-3}$ & 7.26 $\pm$ 0.03 & 0.60 & Y & $-555$ & $-965$ & $-1385$ & $63.4$ \\ 
J0525-2406 & 6.539 & 4.7$\,^{+3.4}_{-1.4}$ & 4.7$\,^{+8.1}_{-2.9}$ & 18$\,^{+7}_{-5}$ & 6.28 $\pm$ 0.05 & 0.83 & Y & $-340$ & $-575$ & $-835$ & $62.1$ \\ 
J2310+1855 & 6.003 & 15.2$\,^{+2.2}_{-1.6}$ & 18.2$\,^{+6.3}_{-4.4}$ & 81$\,^{+33}_{-24}$ & 13.79 $\pm$ 0.11 & 0.83 & Y & $-460$ & $-855$ & $-1275$ & $120.9$ \\ 
P009-10 & 6.004 & 3.6$\,^{+1.8}_{-0.9}$ & 3.6$\,^{+3.9}_{-1.8}$ & 57$\,^{+24}_{-17}$ & 5.75 $\pm$ 0.05 & 1.57 & N & --- & --- & --- & --- \\ 
P083+11 & 6.340 & 6.0$\,^{+1.9}_{-1.2}$ & 6.2$\,^{+4.4}_{-2.4}$ & 35$\,^{+14}_{-10}$ & 7.98 $\pm$ 0.08 & 1.13 & Y & $-415$ & $-905$ & $-1430$ & $66.8$ \\ 
P183+05 & 6.439 & 8.2$\,^{+1.3}_{-0.8}$ & 9.7$\,^{+3.5}_{-2.2}$ & 42$\,^{+17}_{-12}$ & 9.16 $\pm$ 0.06 & 0.94 & Y & $-565$ & $-875$ & $-1205$ & $54.3$ \\ 
P231-20 & 6.586 & 4.7$\,^{+0.9}_{-0.6}$ & 4.9$\,^{+1.9}_{-1.3}$ & 31$\,^{+13}_{-9}$ & 5.86 $\pm$ 0.04 & 0.63 & Y & $-530$ & $-735$ & $-950$ & $26.1$ \\ 
\tableline 
J1319+0950\tablenotemark{b} & 6.133 & 7.6$\,^{+3.1}_{-1.6}$ & 8.6$\,^{+8.1}_{-3.8}$ & 51$\,^{+21}_{-15}$ & 8.56 $\pm$ 0.12 & 0.95 & Y & $-150$ & $-310$ & $-510$ & $14.4$ \\ 
P036+03\tablenotemark{c} & 6.540 & 5.1$\,^{+1.1}_{-0.8}$ & 6.5$\,^{+3.6}_{-2.3}$ & 65$\,^{+27}_{-19}$ & 5.00 $\pm$ 0.05 & 0.71 & N & --- & --- & --- & --- \\ 
J2054-0005\tablenotemark{d} & 6.039 & 7.6$\,^{+1.0}_{-0.7}$ & 11.3$\,^{+4.1}_{-2.6}$ & 32$\,^{+13}_{-9}$ & 5.72 $\pm$ 0.02 & 0.42 & Y & $-670$ & $-1100$ & $-1575$ & $245.2$ \\ 
\enddata 
\tablecomments{The dust size \rdust is the circularized half-width-at-half-maximum of a 2D Gaussian 
  fit to the continuum images. We assume 10\% uncertainties on these values. 
  Uncertainties on \vfifty, \vef, \vmax, and \EWvth are dominated by systematics in the 
  fitting method rather than statistical errors. We estimate 10\%, 15\%, 20\%, and 10\% uncertainties on 
  these values.}\tablenotetext{a}{The luminosities, $S_{\mathrm{119\um}}$, and $r_{\mathrm{119\um}}$ have been corrected for lensing magnification \citep{fan19,yue21}.} 
\tablenotetext{b}{OH data from \citet{herreracamus20}.} 
\tablenotetext{c}{OH data from \citet{butler23}.} 
\tablenotetext{d}{OH data from \citet{salak24}.} 
\end{deluxetable*}

Following the classification scheme of \citetalias{spilker20a}, we consider any source with significant absorption more blueshifted than the \cii emission line to be an unambiguous outflow. This is more restrictive than some other definitions (e.g. $\vfifty < -50$\,\kms; \citealt{rupke05}), but ultimately none of our quasar outflows are `borderline' cases so the exact definition used is of little importance. Unlike the \citetalias{spilker20a} DSFGs, no quasar solely shows OH absorption at systemic velocities, so deep systemic absorption cannot pull velocity metrics toward systemic velocities. By this metric, 6/8 of the quasars we observed show clear outflows; the remaining two quasars (J0305-3150 and P009-10) show OH purely in emission at systemic velocities. Including the additional literature quasars, 8/11 of the $z>6$ quasars with OH observations show evidence for outflows (Sec.~\ref{detection}). No quasar shows evidence of redshifted OH absorption, which would indicate the presence of inflowing molecular gas.

\begin{figure}[t]
\begin{centering}
\includegraphics[width=\columnwidth]{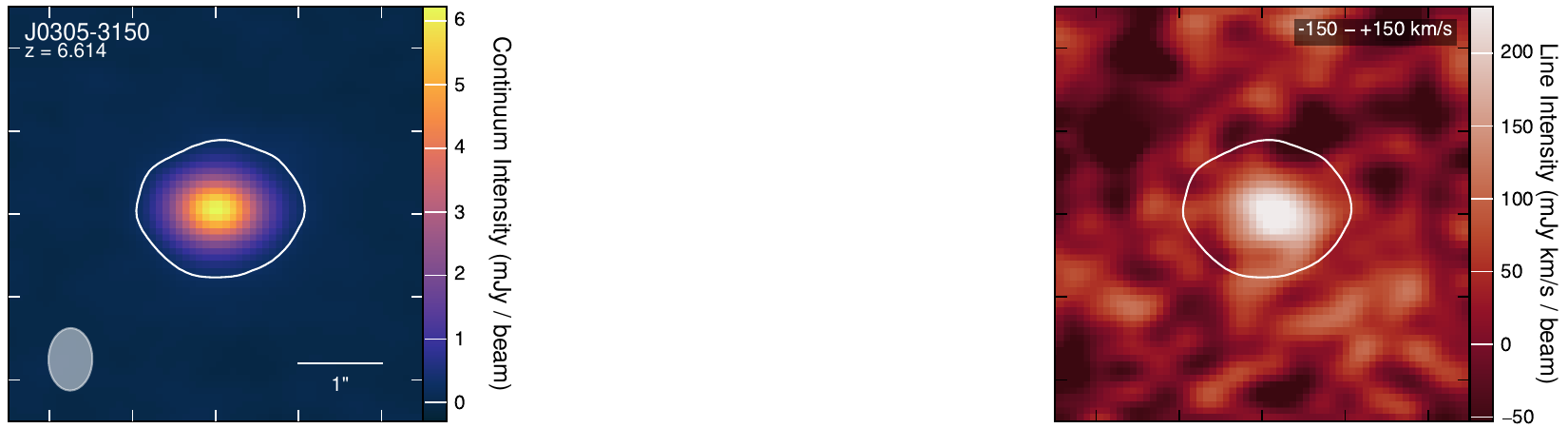}
\includegraphics[width=\columnwidth]{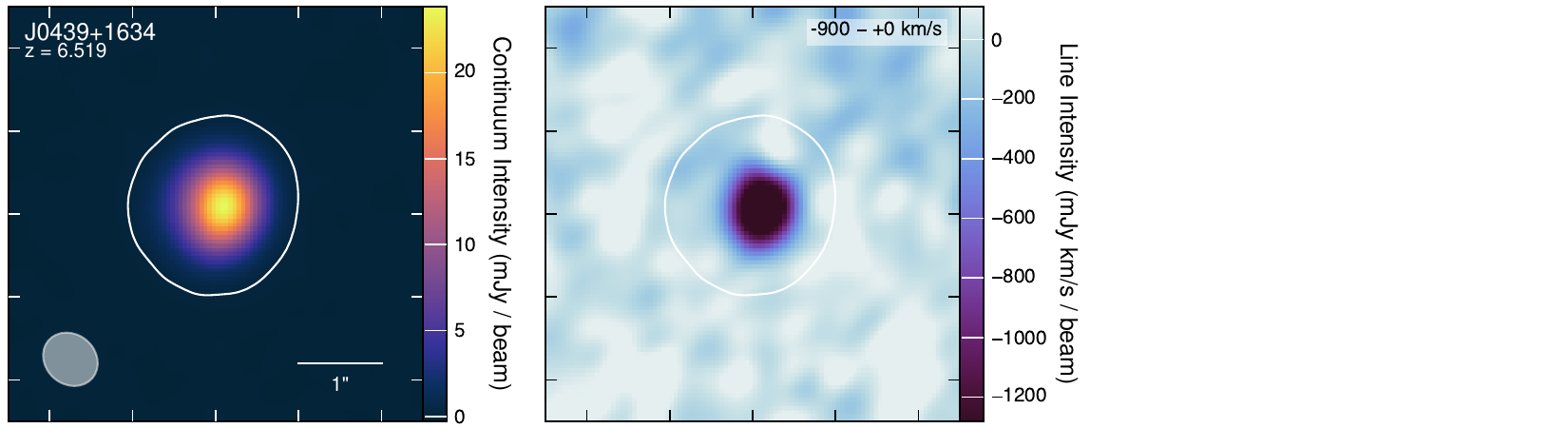}
\includegraphics[width=\columnwidth]{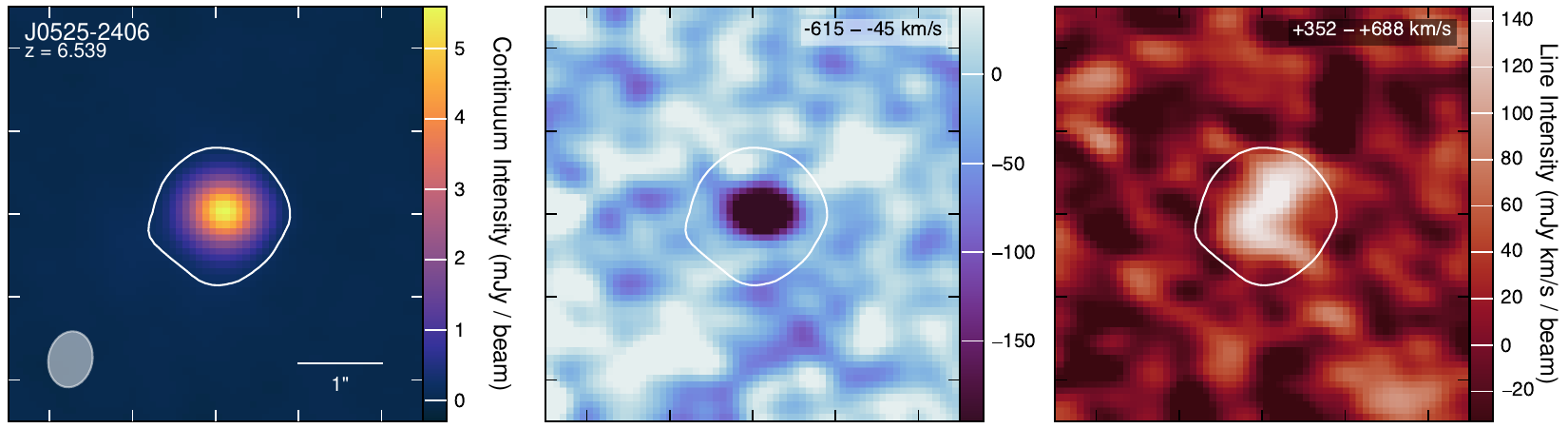}
\includegraphics[width=\columnwidth]{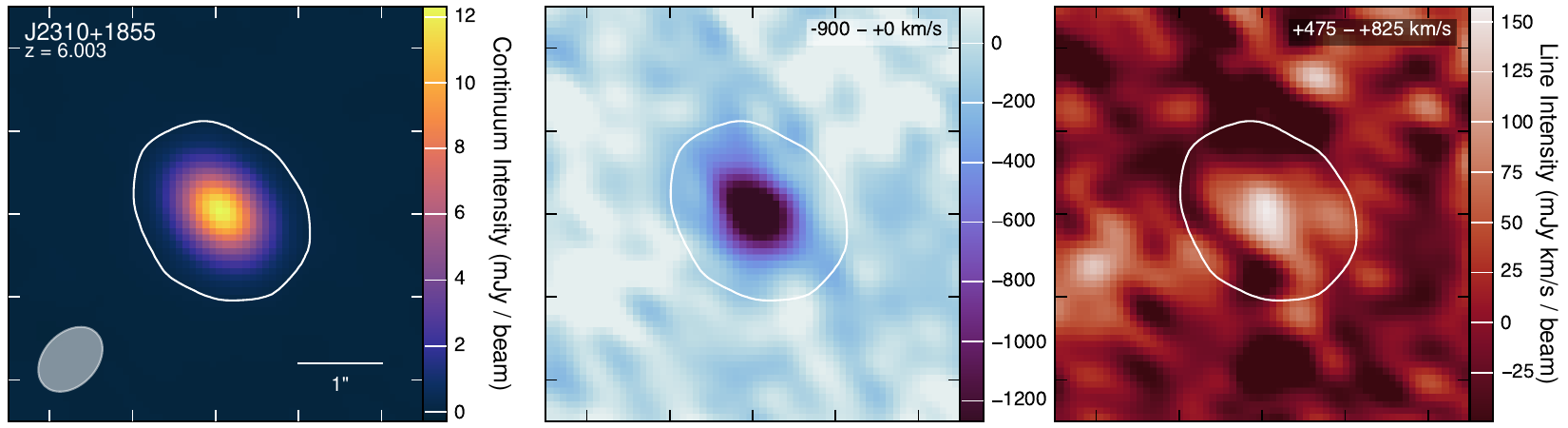}
\includegraphics[width=\columnwidth]{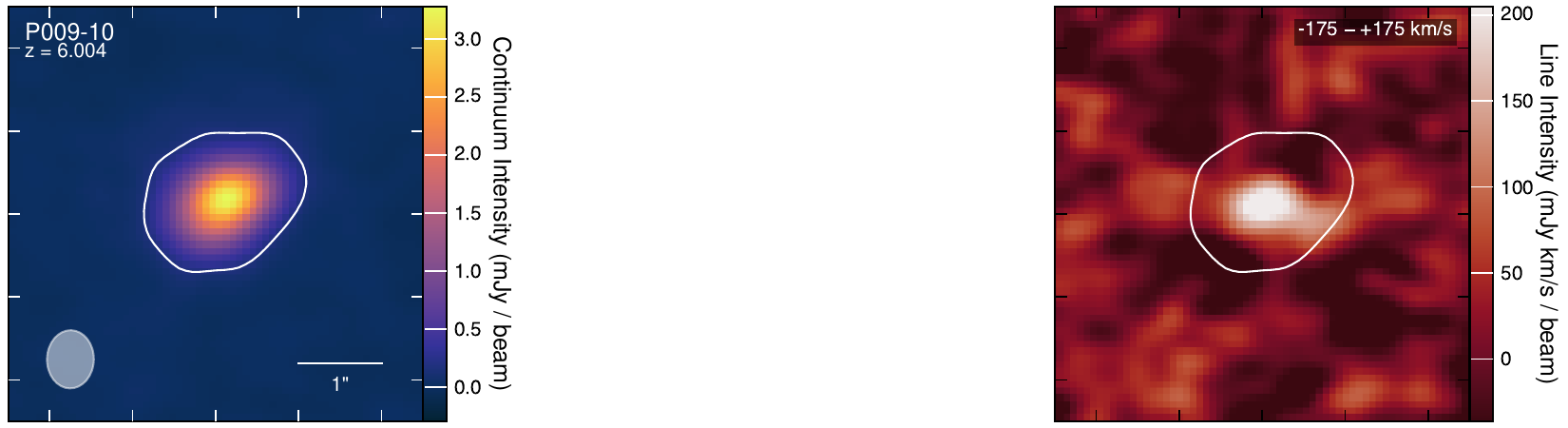}
\includegraphics[width=\columnwidth]{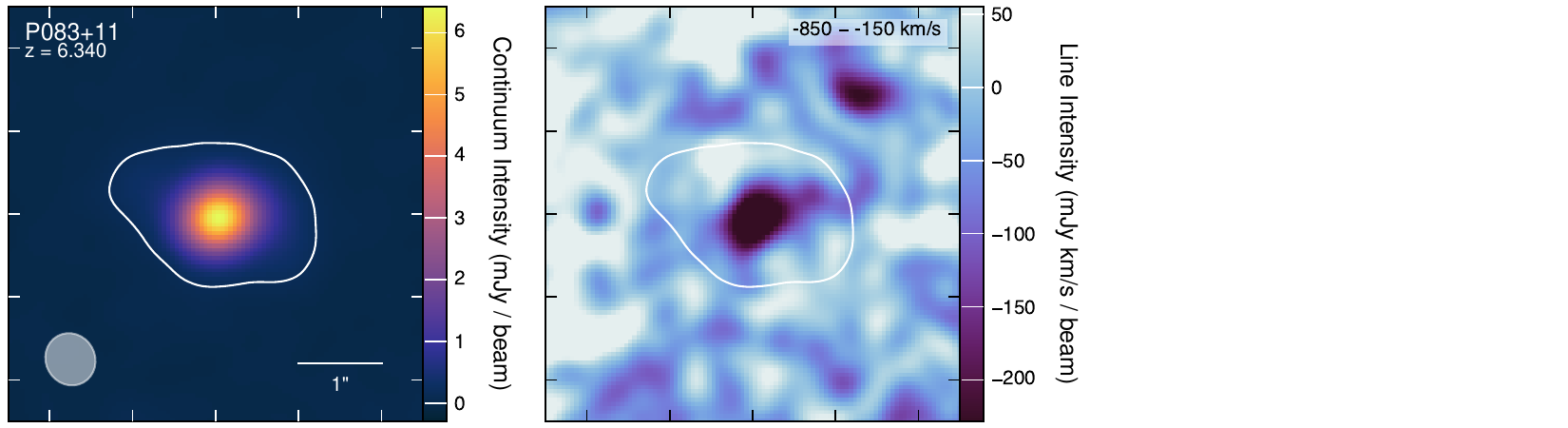}
\includegraphics[width=\columnwidth]{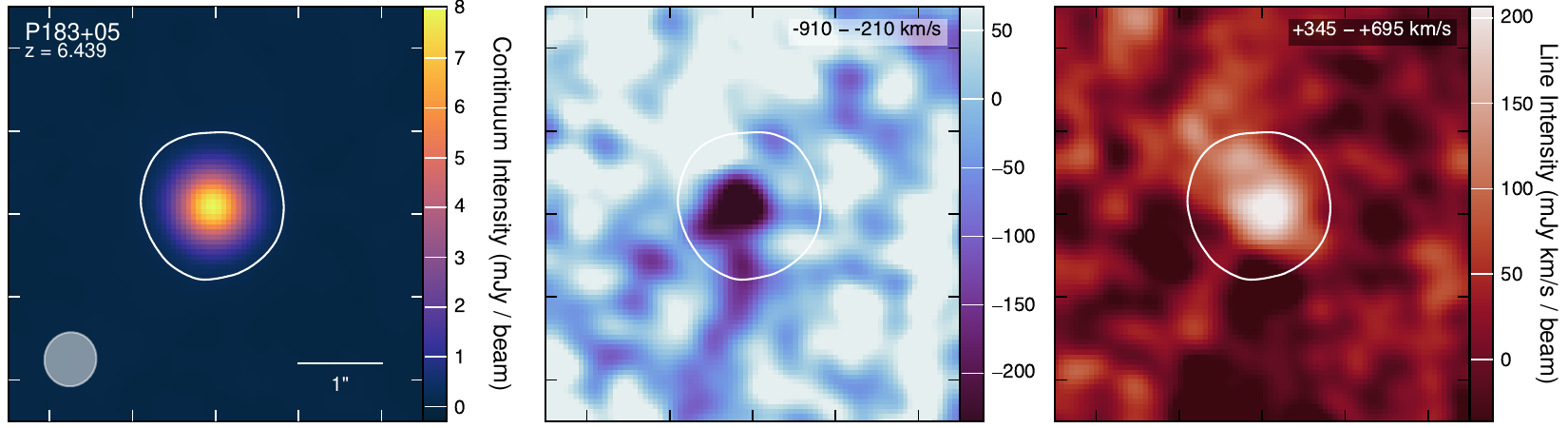}
\includegraphics[width=\columnwidth]{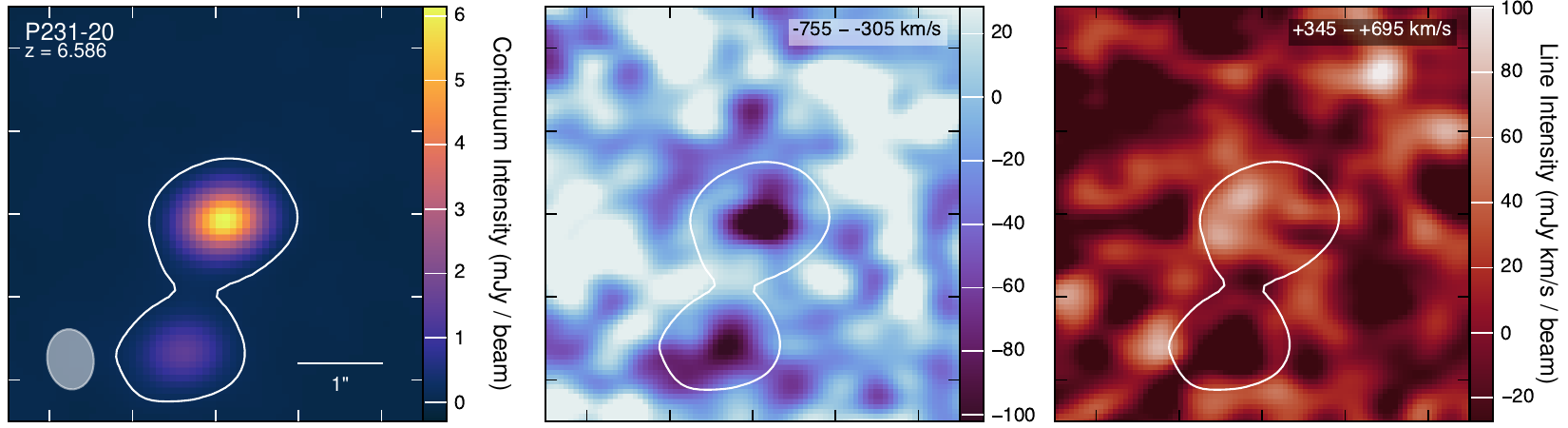}\\
\end{centering}
\caption{
Images of the rest-120\,\um continuum (left column) and OH absorption (center) and emission (right), if present. The line maps are integrated over the velocity ranges noted in each panel, designed to isolate the absorption and emission components as much as possible given the spectral profiles (Fig.~\ref{fig:ohspectra}). Contours show the 10$\sigma$ level of the continuum, and the synthesized beam is shown at lower left. A known nearby companion source is visible in P231-20 \citep{decarli17}.
}\label{fig:images}
\end{figure}

\subsection{Far-IR SED Fitting, SF and AGN Luminosities} \label{sedfitting}

To assess the relative importance of star formation and the AGN in driving outflows, we estimate the luminosities arising from each of these processes. For the AGN luminosity \LAGN, we follow standard practice in the quasar literature and apply a bolometric correction to the measured rest-frame 3000\,\AA ~luminosity $L_{3000}$. We take the bolometric correction from \citet{richards06}, namely $\LAGN = 5.15 L_{3000}$. We calculate \LAGN using the compilation of $L_{3000}$ from \citet{mazzucchelli23}, which has a typical uncertainty of $\approx$0.1--0.2\,dex for our sample quasars and does not include any uncertainty in the bolometric correction factor. For the luminosity arising from star formation, we fit to the available long-wavelength photometry (Sec.~\ref{ancillary}), as described below. For the four quasars with Herschel detections or upper limits in the rest-frame mid-IR, we use these same fits to double-check the UV-based \LAGN estimates. Our SED fits are shown in Appendix~\ref{appsedfits}.

We fit a standard modified blackbody function to the IR photometry. We fix the long-wavelength slope index $\beta = 2.0$ and the wavelength where the dust optical depth is unity $\lambda_0 = 100$\,\um. We join the modified blackbody with a mid-IR powerlaw component on the short-wavelength side of the SED. The two components are merged at the wavelength where the mid-IR powerlaw index equals the derivative of the modified blackbody component \citep[e.g.][]{drew22}. We sample the best-fit SEDs using a Markov Chain Monte Carlo technique and integrate each sample to measure the IR (integrated over rest 8--1000\,\um) and far-IR (40--120\,\um) luminosities \citep[e.g.][]{spilker16}. As is typical in the quasar literature, we assume the modified blackbody component comes from cold dust heated by star formation in the host galaxies, while the powerlaw component represents warm and hot dust heated by the central AGN. A very similar SED fitting procedure was used for the \citetalias{spilker20a} comparison sample and the low-redshift galaxies, both of which have good mid-IR coverage from Herschel and IRAS, respectively.

The number and wavelength coverage of the ancillary measurements varies widely among our sample quasars, which could bias our luminosity estimates. In Appendix~\ref{appsedfits} we test for this effect by refitting the IR SED of each quasar using only rest-frame 160 and 120\,\um continuum measurements, which are available for all sources. We find no evidence for a systematic offset between the IR or FIR luminosities between the fits to the limited vs. full photometry. The luminosities are consistent within the mutual 1$\sigma$ uncertainties, and are equal to less than a factor of 2.

For the quasars with Herschel detections, our SED fits favor a powerlaw index $\alpha \lesssim 1.0$. This is shallower compared to that typically found in non-quasars (for example \citealt{casey12} advocate $\alpha = 2.0$ for most galaxies), a consequence of the greater importance of AGN-heated hot dust in quasars \citep[see also][]{leipski14}. For these quasars, our estimates of the fractional contribution of the AGN to the total IR luminosity \fagn is consistent with literature studies \citep{lyu16,tripodi22,salak24} and the criterion based on the 30\,\um/15\,\um flux ratio \citep{veilleux09}. For the quasars observed but undetected by Herschel/SPIRE, the data constrain $\alpha \gtrsim 0.5$. For simplicity and consistency across the sample, we adopt $\alpha = 1.0$ for all quasars. By varying $\alpha$ in the range 0.5 -- 2.0, we find this choice has a $\lesssim$15\% impact on our estimates of \lfir, while \lir is obviously more severely impacted. For this reason, in the remainder of this work we primarily make use of \lfir when appropriate, which is reasonably well-constrained by the available data.

Assuming the powerlaw component of our IR SED fits arises from AGN-heated dust, we find that the UV-based \LAGN estimates are $2\times$ higher on average than the IR-based \LAGN ($1.8\times$ considering only the two Herschel-detected quasars). This could suggest that the UV bolometric correction factors should be lower to make the UV estimate match the IR, or that some fraction of the modified blackbody component also arises from the AGN to make the IR match the UV. 

We assume the luminosity in the modified blackbody component is the result of star formation heating the cold dust, i.e. $\lir{}_\mathrm{,mbb} \equiv \LSF$. We double-checked our estimates of \LSF for all quasars using the \cii emission as an independent SFR estimate. For the modified blackbody component of our fits, we estimate SFRs as $\mathrm{SFR}_{\mathrm{IR}}/(\Msol/\mathrm{yr}) = 1.4\times10^{-10} \, \lir{}_\mathrm{,mbb}/\Lsol$ for a \citet{chabrier03} initial mass function. We compared these SFRs with those estimated from the \cii luminosity available for all of our sample quasars using the calibration of \citet{delooze14}. We found a mean (median) value $\mathrm{SFR}_{\mathrm{CII}} / \mathrm{SFR}_{\mathrm{IR}} = 1.2$ (0.9), with a maximum disagreement at the factor-of-2 level.

Having cross-checked our estimates of both \LAGN (between rest-UV and IR methodologies) and \LSF (between IR and \cii), we conclude that neither quantity is wildly inaccurate. Our conclusions in the remainder of this work would not be changed if we adopted either the $\approx$2$\times$ lower IR-based \LAGN or the $\approx$1.2$\times$ higher \cii-based SFRs.

\subsection{Outflow Rates, Masses, and other Properties} \label{outflowcalcs}

We calculate outflow properties using a subset of the techniques described in detail in \citetalias{spilker20b}. We assume the same outflow geometry (the `time-averaged thin shell'; \citealt{rupke05}), which relates the outflow rate, mass, momentum, and kinetic power. To summarize, this means:
\begin{subequations} \label{eq:outflow}
\begin{equation}
  \Mout = \Mdot \Rout / \vout,
\end{equation}
\begin{equation}
  \pdot = \Mdot \vout,
\end{equation}
\begin{equation}
  \Edot = \frac{1}{2}\Mdot \vout^2,
\end{equation}
\end{subequations}
with \Mout the mass of the outflow, \Mdot the outflow rate, \Rout the outflow radius, \vout the characteristic velocity, \pdot the momentum rate, and \Edot the outflow kinetic power. For consistency with the literature, we assume $\vout \equiv \vef$, the velocity above which 85\% of the OH absorption takes place (Sec.~\ref{specfitting}).

A much more detailed description of the assumptions, limitations, and multiple techniques used to estimate outflow properties is available in \citetalias{spilker20b}. Here we provide only a brief summary, highlighting differing assumptions made here in comparison to other literature works. We reanalyze and apply the same method to the literature $z>6$ quasars with OH outflows for consistency. We attempt to isolate outflowing gas by measuring the OH equivalent width at velocities more blueshifted than --200\,\kms to limit the impact of any systemic emission/absorption on the outflow properties. Other velocity cuts do not significantly change our outflow property estimates because the blueshifted equivalent width is used only in a totally empirical way. We assume an outflow size $\Rout = \rdust$, the effective radius of the 120\,\um dust emission measured in the ALMA continuum images.

Because OH 119\,\um is known to be highly optically thick even in the line wings \citep[e.g.][]{fischer10,gonzalezalfonso17}, we use empirical estimators of the outflow rates \citepalias[e.g.][]{herreracamus20}. In \citetalias{spilker20b} we calibrated outflow rate estimates using a low-redshift `training sample' of galaxies with OH- or CO-detected outflows assembled from the literature. We repeat this exercise in the same way; the only change is the addition of the newly-available sample of ULIRGs from \citet{lamperti22} to the low-redshift collection. The assumed outflow geometry is the same in all cases. We note that, because this is a simple empirical fit, any deviation from a spherical outflow geometry (i.e. outflow covering factor $<$1) is de facto absorbed into the uncertainties. The low-redshift sample includes galaxies with a wide range of $\fagn = 5-80$\%, so we hope that it is reasonably applicable to our sample of $z>6$ quasars. 

Following \citetalias{herreracamus20}, we fit a relation of the form
\begin{equation}\label{eq:mdot}
\Mdot = m (\EWvth \sqrt{L/10^{12}\Lsol}) + b,
\end{equation}
with the outflow rate in \Msol/yr and \EWvth, the OH 119\,\um equivalent width measured for velocities $v < -200$\,\kms, in \kms. In \citetalias{spilker20b}, $L \equiv \lir$, $m = 1.40^{+0.21}_{-0.25}$, and $b = 180^{+40}_{-30}$. This is primarily an empirical model, but physically asserts that the outflow rate is proportional to the column density (from \EWvth) and the outflow size (from $\sqrt{L}$ via a Stefan-Boltzmann argument), and that the outflows remain self-similar with increasing luminosity. For our quasar sample, \lfir (40--120\,\um) is better constrained than \lir (8--1000\,\um) due to the incomplete Herschel coverage. To understand the impact of this difference on \Mdot, we repeated the same exercise using $L \equiv \lfir$, finding $m = 1.56^{+0.23}_{-0.25}$ and $b = 190 \pm 30$. Both relations reproduce the independently-measured outflow rates of the low-redshift sample equally well.

\begin{figure}
\includegraphics[width=\columnwidth]{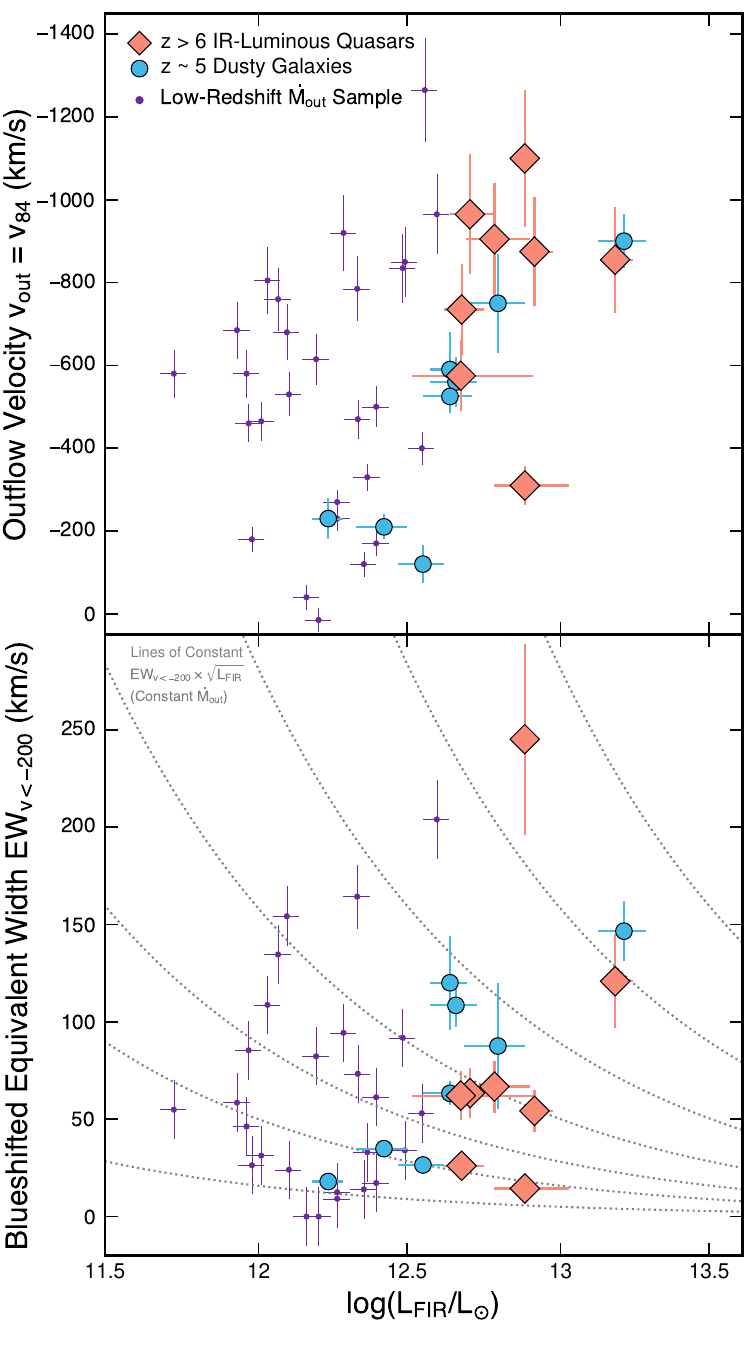}
\caption{
Comparison of the observables we use to estimate the physical properties of molecular outflows based on the sample of low-redshift objects with OH 119\,\um detections and published \Mdot. The upper panel highlights that the OH outflows in the high-redshift quasars are typically faster than at low redshift, which is most relevant for estimating \pdot and \Edot. The lower panel directly compares \EWvth and \lfir, the two quantities we combine to estimate \Mdot. Dotted gray lines illustrate lines of constant \Mdot under the assumed parameterization.
}\label{fig:voutew200}
\end{figure}

Figure~\ref{fig:voutew200} compares these galaxy samples in the quantities relevant for the outflow property calculations. We notice that the high-redshift quasars are more luminous than the other samples (as expected by our selection), have faster outflow velocities, and/but relatively low blueshifted equivalent widths. We explore the significance of these points in Sections~\ref{velocities}~and~\ref{structure}. The lower panel of Fig.~\ref{fig:voutew200} shows lines of constant $\EWvth \sqrt{\lfir}$, which Equation~\ref{eq:mdot} asserts to be lines of constant \Mdot. While our technique does rely, to some extent, on extrapolation from the low-redshift samples (especially in luminosity), this extrapolation is not terribly severe in combination with the equivalent widths.

We applied both versions of Eq.~\ref{eq:mdot} to our quasar sample and the non-quasar comparison sample \citepalias{spilker20a}. For the non-quasar DSFGs, both methods produced very similar outflow rates due to the uniformly excellent coverage of the far-IR SEDs. Because the $z>6$ quasars have much more power in the hot dust component, the \lir-based method results in $\approx$30\% higher outflow rates than the \lfir-based method on average. In the remainder of our analysis, we simply average the results from the two methods. Improved estimates of \Mdot will require observations of additional OH transitions, alternate outflow tracer species, and/or high resolution to lessen the uncertainties due to the OH abundance, optical depth, and geometric/projection effects. Outflow molecular masses, momenta, and kinetic powers are derived using the assumed outflow geometry. Table~\ref{tab:outprops} summarizes these outflow properties.


\begin{deluxetable*}{lcccc} 
\tablecaption{Quasar outflow properties \label{tab:outprops}}
\tablehead{
\colhead{Source} & 
\colhead{\Mdot} & 
\colhead{\Mout} & 
\colhead{\pdot} & 
\colhead{\Edot} \\ 
\colhead{} & 
\colhead{\Msol/yr} & 
\colhead{10$^8$\,\Msol} & 
\colhead{10$^5$\,\Msol/yr \kms} & 
\colhead{10$^{10}$\,\Lsol} 
} 
\startdata 
J0439+1634 & 470 $\pm$ 200 & 6.0$\,^{+4.0}_{-3.0}$ & 4.5$\,^{+2.9}_{-2.3}$ & 3.6$\,^{+3.2}_{-2.1}$ \\ 
J0525-2406 & 450 $\pm$ 240 & 6.3$\,^{+5.1}_{-3.7}$ & 2.6$\,^{+2.0}_{-1.5}$ & 1.2$\,^{+1.3}_{-0.8}$ \\ 
J2310+1855 & 1150 $\pm$ 320 & 10.9$\,^{+5.5}_{-4.1}$ & 9.8$\,^{+4.6}_{-3.8}$ & 6.9$\,^{+4.8}_{-3.3}$ \\ 
P083+11 & 510 $\pm$ 220 & 6.2$\,^{+4.2}_{-3.1}$ & 4.6$\,^{+3.0}_{-2.4}$ & 3.4$\,^{+3.0}_{-2.0}$ \\ 
P183+05 & 500 $\pm$ 200 & 5.3$\,^{+3.4}_{-2.5}$ & 4.4$\,^{+2.7}_{-2.1}$ & 3.1$\,^{+2.7}_{-1.8}$ \\ 
P231-20 & 300 $\pm$ 180 & 2.5$\,^{+2.2}_{-1.7}$ & 2.2$\,^{+1.9}_{-1.5}$ & 1.3$\,^{+1.5}_{-1.0}$ \\ 
\tableline 
J1319+0950\tablenotemark{a} & 260 $\pm$ 190 & 7.9$\,^{+8.0}_{-6.0}$ & 0.8$\,^{+0.8}_{-0.6}$ & 0.2$\,^{+0.3}_{-0.2}$ \\ 
J2054-0005\tablenotemark{b} & 1610 $\pm$ 420 & 6.0$\,^{+2.9}_{-2.2}$ & 17.7$\,^{+7.9}_{-6.6}$ & 16.0$\,^{+10.6}_{-7.5}$ \\ 
\enddata 
\tablecomments{Uncertainties are derived by propagating the uncertainty in \Mdot (Eq.~\ref{eq:mdot}) 
  through the other outflow properties (Eq.~\ref{eq:outflow}).} 
\tablenotetext{a}{OH data from \citet{herreracamus20}; we re-estimated the outflow properties using our methodology.} 
\tablenotetext{b}{OH data from \citet{salak24}; we re-estimated the outflow properties using our methodology.} 
\end{deluxetable*}

\section{Results and Discussion} \label{results}

\subsection{High molecular outflow detection rate} \label{detection}

\begin{figure}
\includegraphics[width=\columnwidth]{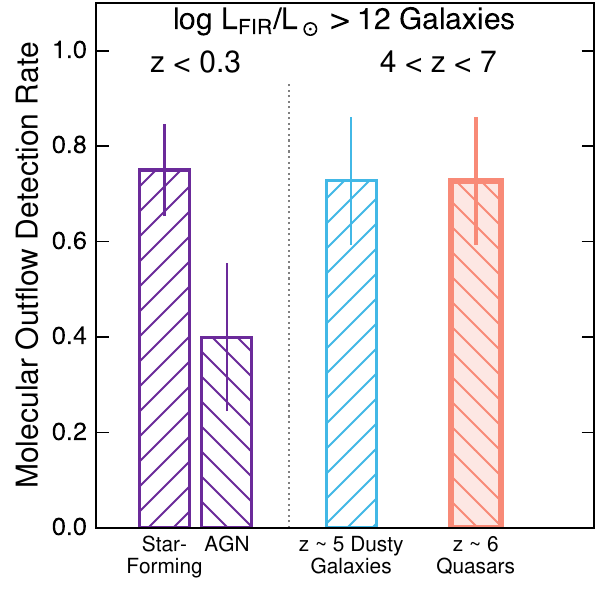}
\caption{
IR-luminous quasars at $z>6$ very commonly show clear evidence of molecular outflows, similar to both $z\sim5$ non-quasar DSFGs and low-redshift starburst-dominated galaxies but somewhat higher than low-redshift AGN-dominated sources. These detection rates are lower limits to the true fraction of sources with molecular outflows, because some outflows may not be detectable in absorption along the line of sight. Uncertainties are from binomial statistics. 
}\label{fig:detrates}
\end{figure}

Our first, most basic result is that cold molecular outflows appear to be very common in the most IR-luminous $z>6$ quasars. Including the literature quasars, we identify strong blueshifted OH absorption in 8/11 objects observed, clear evidence of outflowing gas. The outflow detection rate is 73\% $\pm$ 13\%, assuming binomial statistics. This is a lower limit to the true occurrence rate of cold outflows, because some sources may have outflows in the plane of the sky rather than along the line of sight. We note that our result here is not obviously limited by the depth of the observations: OH was detected (either in emission, absorption, or both) in all quasars, though it is possible that weak OH absorption would not be detectable.

Figure~\ref{fig:detrates} compares the outflow detection rate in our sample to comparison objects at both low and high redshifts. The low-redshift sample is assembled from the superset of sources observed in OH by the Herschel Space Observatory \citep{spoon13,veilleux13,calderon16,stone16,herreracamus20}. We restrict ourselves to objects with $\log \lfir / \Lsol > 12$ to make a more fair comparison with our own sample. A true luminosity-matched comparison is not possible because objects as luminous as those observed at high redshift are exceedingly rare locally.

At low redshifts, the most AGN-dominated galaxies have unambiguous molecular outflows less frequently than starburst-dominated objects. This may be due to a geometric effect: while starburst-driven outflows are preferentially perpendicular to the disc, AGN-driven winds may have more random inclinations \citep[e.g.][]{veilleux13,stone16,lamperti22}. We see no such difference at high redshift: both our sample of IR-luminous $z>6$ quasars and the \citetalias{spilker20a} $z\sim5$ DSFGs have $\sim$70\% outflow detection rates.\footnote{Coincidentally both the detection rate and uncertainty are identical between the quasars and DSFGs, because the number of outflows and total sample sizes are the same.}

We can interpret these similar detection rates in multiple ways. First, it could suggest that the molecular outflows are completely unrelated to the presence of the quasars. If the outflows are driven primarily by host star formation in both quasars and non-quasars at high redshift, the detection rate would be expected to be the same. However, as we show in the next subsections, we find some systematic differences between the outflows in the quasar vs. non-quasar DSFG samples. If host galaxy star formation primarily drives these outflows, it is clear that the additional energy and momentum from the quasars help `boost' the outflows to be faster and more powerful, a point we return to in Sec.~\ref{energetics}.

Alternatively, it could suggest that the outflows in the DSFGs at $z\sim5$ are actually primarily driven by AGN rather than star formation (essentially the converse of the argument in the previous paragraph). This seems unlikely, both because there are clear differences in the outflow properties between samples and because there is presently no evidence for AGN in the $z\sim5$ DSFGs and the outflow energetics do not require an AGN in any object \citepalias{spilker20b}. The lack of AGN is based on the faint rest-frame mid-IR continuum \citepalias{spilker20a}, which is much better constrained for the DSFGs than our quasars. The lack of an AGN is confirmed by recent JWST spectroscopy in the rest-optical \citep{birkin23} and mid-IR \citep{spilker23} for one of the DSFGs. 

Finally, it is possible that the molecular outflows in our $z>6$ quasars, unlike at lower redshifts, are not yet beginning to subside even though a low-obscuration line of sight directly to the AGN has been cleared. This could occur if the outflows in these quasars are widespread throughout the galaxies, impacting a larger region than the path cleared to the central engine. This explanation also allows for different outflow properties between the high-redshift quasars and non-quasars because the driving mechanism is not the same, as we discuss further below.

\subsection{High-velocity molecular outflows in $z>6$ quasars} \label{velocities}

\begin{figure*}
\centering
\includegraphics[width=0.45\textwidth]{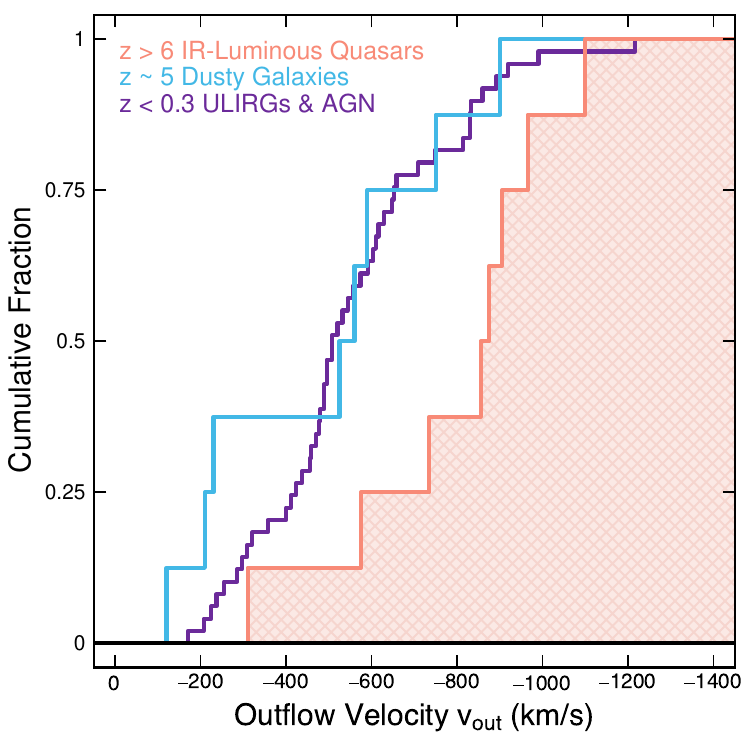}
\includegraphics[width=0.45\textwidth]{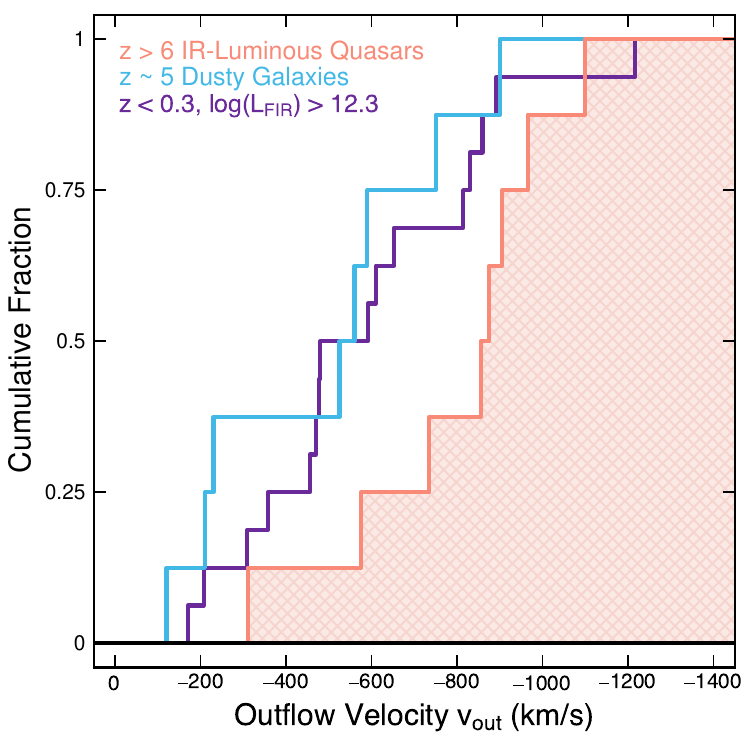}
\caption{
Molecular outflows in $z>6$ quasars are significantly faster than those typically seen in low-redshift ULIRGs and AGN or $z\sim5$ non-quasar DSFGs, by several hundred \kms on average. This difference remains if we restrict the low-redshift literature sample to the most luminous objects (though still not a matched-luminosity subset; see Fig.~\ref{fig:voutlagn}). In this Figure, only objects said to exhibit molecular outflows by the original authors are shown. All galaxies in this Figure were detected in OH 119\,\um absorption, and we plot the same $\vout \equiv \vef$ velocity metric for all samples.
}\label{fig:velcdf}
\end{figure*}

\begin{figure*}
\centering
\includegraphics[width=0.9\textwidth]{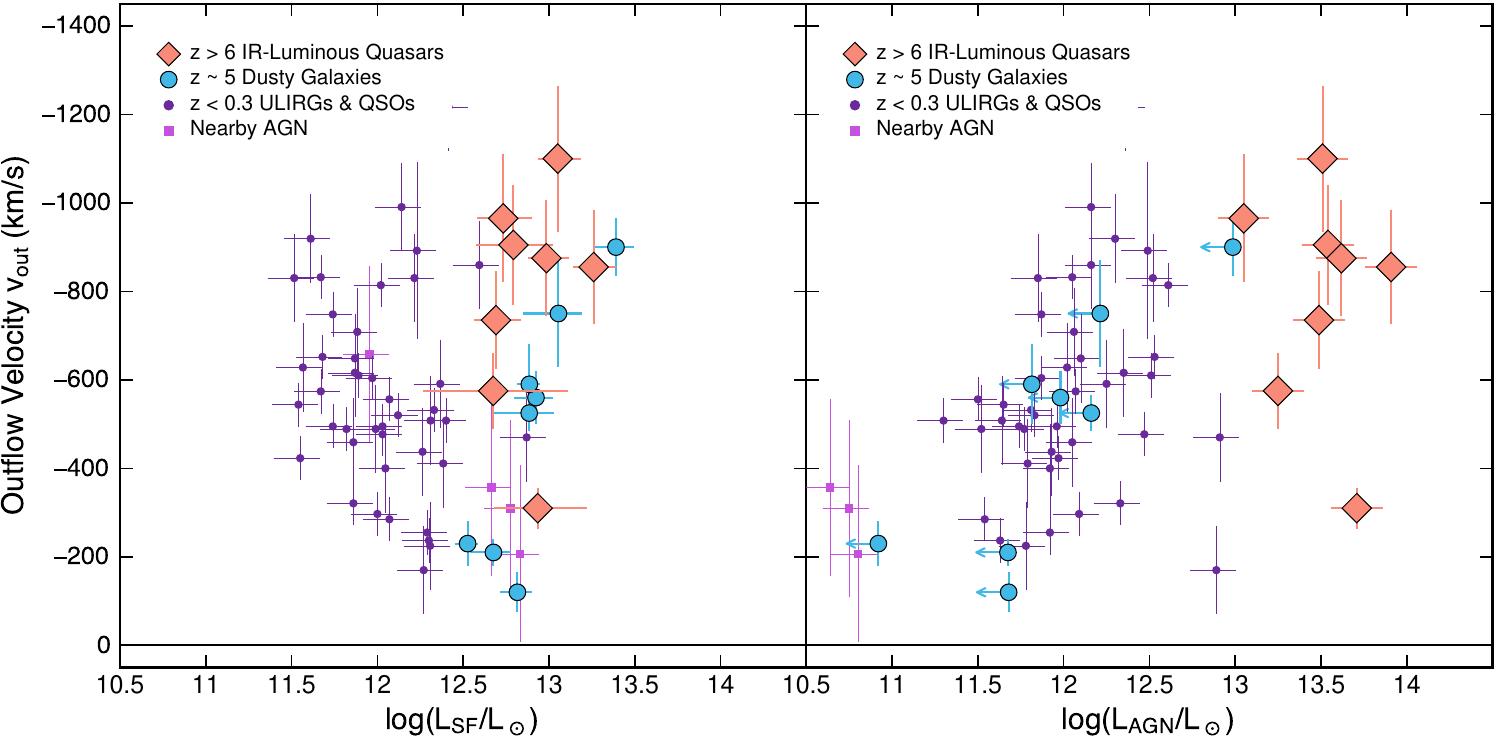}
\caption{
When considering the luminosity from star formation and AGN separately in the low- and high-redshift sources with OH 119\,\um absorption, we find a clear trend between \vout and \LAGN but no similar trend with \LSF. Due to the large scatter, the trend with \LAGN emerges thanks to the large dynamic range probed by these samples; it is thus not clear whether a similar trend could exist for \LSF. 
}\label{fig:voutlagn}
\end{figure*}

We turn our attention to another basic observable property of the OH spectra, the velocity of the blueshifted absorption. It is clear from Fig.~\ref{fig:ohspectra} that most of the quasars have obvious fast outflows. We quantify this in terms of the common outflow velocity metric $\vout \equiv \vef$, the velocity above which 84\% of the absorption takes place. We find similar results using other outflow velocity metrics. Figure~\ref{fig:velcdf} shows the cumulative distributions of \vef for our high-z quasars and the comparison high-z non-quasars and low-redshift Herschel OH superset.\footnote{Figures~\ref{fig:velcdf} and \ref{fig:voutlagn} have a larger number of low-redshift comparison galaxies because they do not require published outflow \textit{physical} properties (e.g. \Mdot), unlike Fig.~\ref{fig:voutew200} and later plots.}

It is strikingly apparent that the $z>6$ quasars have, on average, substantially faster outflow velocities than either comparison sample. A two-sided KS test returns probabilities that the \vout distribution of the high-z quasars is drawn from the same distribution as either the DSFG or low-redshift samples of $p < 0.05$. The median outflow velocity for both the $z\sim5$ DSFGs and the low-redshift assortment of ULIRGs and QSOs is $\approx$550\,\kms, while the median for the $z>6$ quasars is nearly 900\,\kms. Even the minimum (least blueshifted) outflow velocity for the high-z quasars is about the same as the median for the other samples. 

The difference in these distributions remains even if we restrict the low-redshift comparison sample to the most luminous objects with $\log \lfir/\Lsol > 12.3$ (as before, a true luminosity-matched comparison is not possible due to the lack of sufficiently extreme low-redshift objects). The right panel of Fig.~\ref{fig:velcdf} restricts the low-redshift sample, and while the median outflow velocity increases to $\approx$600\,\kms, it is still far below what we observe in the high-redshift quasars.

Figure~\ref{fig:voutlagn} shows the outflow velocities of each source as a function of the luminosity arising from star formation and the AGN separately, which we estimated previously (Sec.~\ref{sedfitting}). Past studies at low redshift found no signs of a correlation between \vout and \LSF but a mild trend with \LAGN. Our analysis of high-redshift quasars and non-quasars confirms and reinforces this picture, now over 4 orders of magnitude in \LAGN. This conclusion would also hold true if we used our (extremely uncertain for most sources) IR-based estimates of \LAGN. Although we see a trend between \vout and \LAGN, it is difficult to conclude that there is \textit{not} a trend between \vout and \LSF, because the dynamic range in \LSF probed by the available data is much smaller than in \LAGN.

The scatter in \vout in Figure~\ref{fig:voutlagn} is very large at a given \LAGN; the trend apparent in this Figure only emerges thanks to the large range in \LAGN covered by the combined low- and high-redshift samples. This scatter probably arises from many sources, chief among them the geometry/inclination of the outflows, the coupling between the AGN luminosity and the outflowing gas, and the underlying contribution from star formation. There is also a large mismatch between the typical timescale for AGN variability and the longer characteristic outflow timescale $\tout = \Rout/\vout \sim$ few Myr (for typical $\Rout \sim$ few kpc and $\vout \sim 1000$\,\kms). High-redshift samples that probe a larger range in \LAGN (and \LSF) will be required to understand the connection with \vout in more detail.

\subsection{The small-scale structure of high-redshift molecular outflows} \label{structure}

At the depth and resolution of our data, none of the blueshifted OH absorption components are spatially resolved. Nevertheless, we can comment on the small-scale structure of these outflows because the line opacity in OH 119\,\um is very high. Essentially 100\% of the continuum light is absorbed along a line of sight that impacts even low column densities of OH molecules, even into the line wings \citep[e.g.][]{fischer10}, so the fractional depth of the OH absorption can be used to estimate the fraction of the background source covered by the outflows (as seen from our perspective). Because we have measured the size of the background continuum emission, we can estimate the effective size of the outflows.

\begin{figure}
\includegraphics[width=\columnwidth]{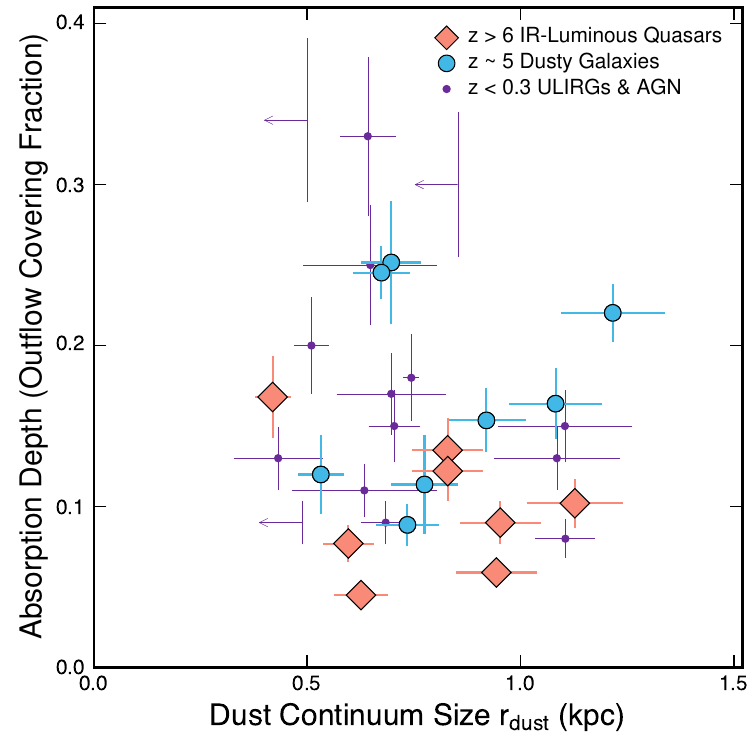}
\caption{
The OH 119\,\um absorption is typically weaker in our $z>6$ quasar sample than the comparison samples. Because OH 119\,\um is very optically thick, the absorption depth directly translates to the outflow covering fraction, implying that the high-redshift quasar outflows are confined to somewhat smaller regions despite their faster velocities. The y-axis shows the peak fractional absorption depth at blueshifted velocities $v < -200$\,\kms, a proxy for the covering fraction of the outflowing gas.
}\label{fig:fcov}
\end{figure}

Figure~\ref{fig:fcov} shows the outflow covering fraction, estimated from the peak fractional depth of the blueshifted OH absorption, as a function of the effective size of the dust continuum emission \rdust. All dust sizes were estimated in a similar way, and are corrected for gravitational lensing where needed (i.e. the DSFGs and J0439+1634). The high-redshift samples use the ALMA in-band $\approx$120\,\um continuum sizes, while the low-redshift sizes are from Herschel/PACS 100\,\um imaging \citep{lutz16}. The peak absorption depths are measured for blueshifted velocities $v < -200$\,\kms, but we find qualitatively similar results for other metrics. The absorption depths are estimated from our modeled OH spectra including only one of the doublet components to avoid double-counting overlapping absorption.

One caveat to this approach is that we are ignoring any dust contained within the outflows, which would `fill in' some of the absorption signal and make our covering fraction estimates smaller than they truly are. Outflows typically contain $\lesssim$10\% of the total galaxy dust mass \citep[e.g.][]{melendez15,barcosmunoz18}, so the dust in the outflow would have to be significantly warmer than the dust in the galaxy to fill in a large part of the absorption profile. 

Ignoring any emission infilling the absorption profiles, Fig.~\ref{fig:fcov} shows that, perhaps surprisingly, the $z>6$ quasar outflows have smaller covering fractions than either the low-redshift or $z\sim5$ DSFG comparison samples. Even though all samples span a comparable range in \rdust, Fig.~\ref{fig:fcov} implies that the quasar outflows are typically confined to smaller regions of their host galaxies. Combining the dust size and covering fraction, we can estimate the size the outflows would have if they were circular. We find a median effective size of the quasar outflows of $\approx$250\,pc, $\approx$30\% smaller than for either the DSFGs or low-redshift sample.

Without resolving the structure of these outflows directly, the interpretation of Fig.~\ref{fig:fcov} must remain unclear. For a simple biconical outflow geometry, one option is that the quasar outflows are more collimated or have a smaller opening angle than the other samples, resulting in smaller covering fractions. This idea has secondary appeal, in that it could also explain the faster outflow velocities if the narrower outflows are fortuitously aligned, for example along the line of sight already cleared to the quasar nucleus. If the outflows are clumpy, like those in the $z\sim5$ DSFGs \citepalias{spilker20a}, the quasar outflows could have either fewer clumps or smaller clumps, on average. 

One $z>6$ quasar with high-resolution data, J2054-0005, was studied in detail by \citet{salak24}. From $\approx$0.2\arc imaging, those authors found OH absorption resolved over a $\gtrsim1$\,kpc region. The resolution was too poor to resolve any small-scale clumpiness. There was no evidence for faster blueshifted velocities toward the center of the galaxy, as expected for a simple expanding spherical outflow, but the slight velocity gradient observed in the absorption is consistent with an inclined biconical geometry. Given the brightness of this and our own quasar sample and the observed absorption depths, we suggest that high-resolution followup observations could be a fruitful avenue toward understanding the structure of high-redshift outflows.

\subsection{Quasar molecular outflow rates are modest} \label{mdot}

We now turn our attention to the physical properties of the molecular outflows, beginning with the cold outflow rates \Mdot. Although these properties are highly uncertain, our goal is first and foremost to compare the $z>6$ quasars with other literature samples on an empirical basis. We compare to the $z\sim5$ DSFGs (with \Mdot estimated the same as for our sample; \citetalias{spilker20b}). At low redshift, we compare to galaxies with outflows detected in OH (primarily ULIRGs and QSOs; \citealt{veilleux13,gonzalezalfonso17,lamperti22}) and galaxies with outflows detected in CO (extending to less-extreme objects; \citealt{lutz20,lamperti22}). Between the two methods, the former are more directly comparable to the high-redshift samples because at least the observable tracer is the same.

\begin{figure*}
\centering
\includegraphics[width=0.95\textwidth]{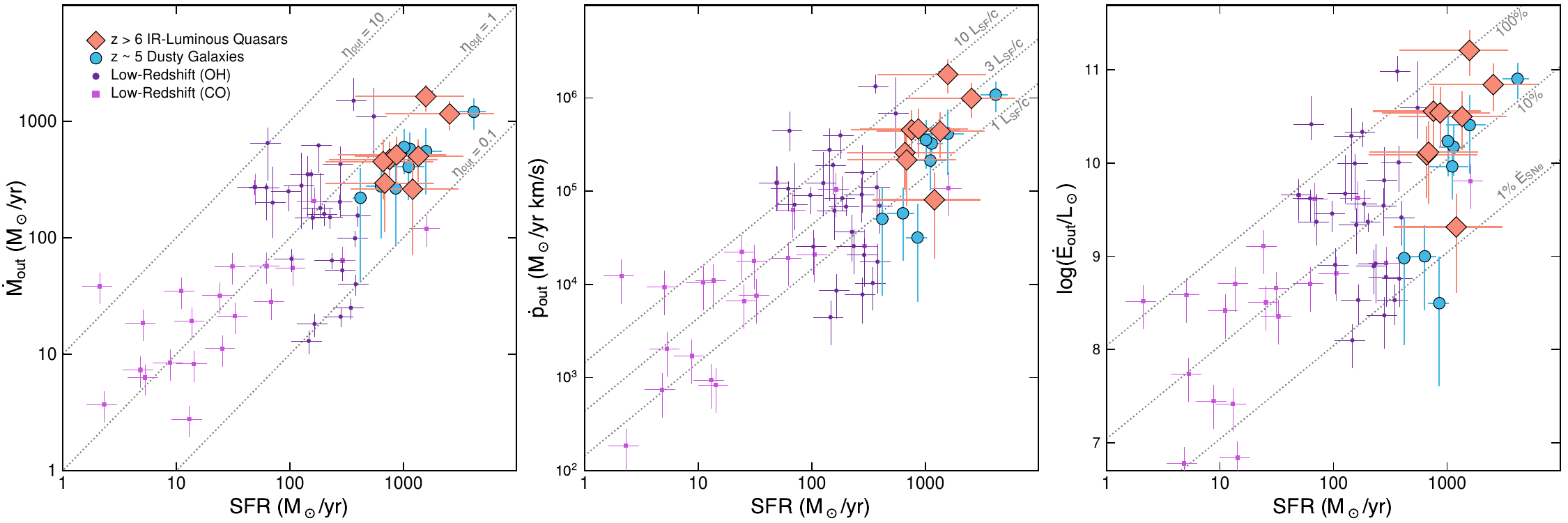}
\includegraphics[width=0.95\textwidth]{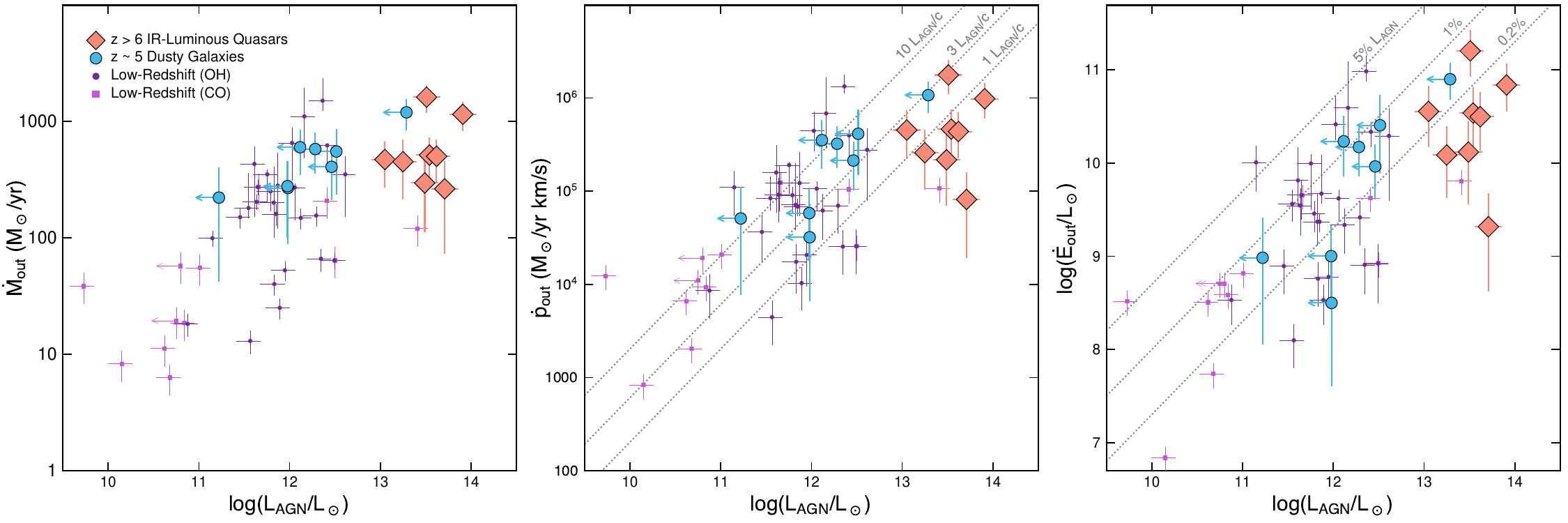}
\caption{
Molecular outflow rates (left column), momentum rates (center), and kinetic power (right) as a function of SFR (top row) or \LAGN (bottom). We find that (i) the $z>6$ quasar outflows have modest mass-loading factors $\eta \sim 0.5$ in the molecular phase (top left panel, Sec.~\ref{mdot}); (ii) the outflows are consistent with being momentum-driven by either supernovae or the AGN with modest momentum boosts $\pdot / (L/c) \lesssim 3$ (center column, Sec.~\ref{energetics}); and/but (iii) power from the AGN is probably needed to explain the kinetic energy in the outflows absent unrealistically high coupling of the available supernova energy to the cold molecular outflow (right column, Sec.~\ref{energetics}).
}\label{fig:outprops}
\end{figure*}

The left panels of Figure~\ref{fig:outprops} show the estimated molecular outflow rates as a function of SFR and \LAGN. The molecular outflow rates are unsurprisingly high, $\Mdot \approx 300-1500$\,\Msol/yr. These are not actually very extreme, however: they are comparable to those found for the similar-luminosity $z\sim5$ DSFGs, and similar to outflow rates found in some low-redshift galaxies with SFRs lower by an order of magnitude. The implied mass loading factors are similarly not especially extreme; the median is $\etaout = \Mdot/\mathrm{SFR} = 0.5$, ranging from 0.2--1. Even if we have significantly overestimated the SFRs (for example if AGN-heated dust contributes significantly to the long-wavelength IR photometry), the mass loading factors would still be quite modest. 

We caution here that there is some circularity in our comparisons. The outflow rates depend on the total $\sqrt{\lir}$ (Sec.~\ref{outflowcalcs}), but we also use IR-based SFRs after subtracting the AGN contribution. We attempted to mitigate this circularity as best as we are able, and cross-checked our results with multiple methods (Sec.~\ref{sedfitting}). First, as is standard in the quasar literature, we estimate the AGN luminosity from the rest-UV continuum, so the outflow properties are independent of \LAGN. Second, we double-checked the AGN-subtracted IR-based SFRs using the \cii luminosity, finding good agreement; SFR$_{\mathrm{CII}}$ is also independent of the outflow rates. It is clear from Fig.~\ref{fig:outprops} that \Mdot and SFR are not totally circular, because we do not find that $\Mdot \propto \mathrm{SFR}^{1/2}$, as Eq.~\ref{eq:mdot} would assert. However, it is also clear that more robust, independent outflow property estimates are sorely needed for our own and other samples.

The similar values for \Mdot between the $z>6$ quasars and $z\sim5$ DSFGs may seem unexpected given the substantially faster outflow velocities we observe in the quasar sample (Fig.~\ref{fig:velcdf}). The solution can be seen in Fig.~\ref{fig:voutew200}: although the quasars are modestly more FIR-luminous than the DSFGs, the blueshifted OH absorption is actually modestly \textit{weaker} in equivalent width. In our estimates for \Mdot, these differences offset, leading to basically identical median $\Mdot \approx 500$\,\Msol/yr. The weaker absorption, as discussed in Sec.~\ref{structure}, most likely results from a smaller wind covering fraction in the quasar sample, either via more collimated outflows or differences in the small-scale structure.

The range in \etaout across the combined high-redshift DSFG and quasar samples is rather narrow, at least in comparison to the nearly 2\,dex spread seen at lower redshift. This could be partially due to limitations in sensitivity; although OH is detected in all sources of both samples, weak blueshifted wings can be difficult to discern on top of the complex absorption/emission doublet profiles. It is also possible that our method to estimate \Mdot (Sec.~\ref{outflowcalcs}) has forced \etaout to take on a narrow range of values because we only use a small number of parameters to link observables to \Mdot. Alternatively, it may simply be that both high-redshift samples span only a very limited range in $\lfir$, $\approx$0.5\,dex (Fig.~\ref{fig:voutew200}). It seems clear that both a broader sample of high-redshift galaxies and improved molecular outflow rates will be needed to understand the connection between star formation and outflow launching.

\subsection{AGN-boosted, momentum-driven outflows} \label{energetics}

In contrast to their similar molecular outflow rates, we expect the $z>6$ quasars to have higher outflow momentum flux ($\pdot = \Mdot \vout$) and kinetic power ($\Edot = 0.5 \Mdot \vout^2$) than the $z\sim5$ DSFGs because these quantities depend more steeply on \vout. We confirm this to be the case, with \pdot (\Edot) typically 1.6 (2.7) times higher in the high-redshift quasars, exactly as expected given the 1.6 times higher outflow velocities. We reiterate that there is some circularity in our comparisons of \pdot and \Edot with the SFR (Sec.~\ref{mdot}), but this should impact \pdot and \Edot less than \Mdot itself as \vout becomes the main controlling parameter. We also recall that \LAGN and \Edot are estimated from different wavelength regimes, and are truly independent.

Outflows driven by winds shocked on the cold gas in the host galaxy can be thought of as `energy driven' if radiative losses are negligible during their expansion, and `momentum driven' if radiative cooling is significant. In the absence of very high dust opacity in the IR, one signature of the former is that the outflow momentum flux can exceed the radiative momentum flux $L/c$ by a large factor, 10 or more, as the outflowing gas expands adiabatically. In momentum-driven winds, however, this momentum boost above the radiative momentum is typically up to a few, $\approx 2-4$ depending on details of the star formation history, initial mass function, and/or coupling of the inner AGN wind and radiation to the larger-scale outflow \citep[see recent reviews by][]{veilleux20,thompson24}. These greater-than-unity momentum boosts do not distinguish the ultimate outflow driving source, as they can be achieved by both supernovae and AGN coupling to the cold gas \citep[e.g.][]{kim15,king15}.

Alternatively, outflows may be driven by the radiation pressure on dust instead of mechanical shock forces. This scenario does permit modest momentum boosts above the radiative momentum flux $L/c$ if the optical depths are so large that trapped photons scatter multiple times \citep[e.g.][]{thompson15,costa18,ishibashi18}, which requires high column densities surrounding the injection region. Compared to energy-driven flows, winds driven by radiation pressure experience milder accelerations, potentially allowing a more substantial cold component to survive and/or reform by cooling \citep{costa18}. Analytic calculations predict scalings $\Mdot \propto L^{1/2}$ and $\Edot \propto L^{3/2}$ in this case \citep{ishibashi18}.

The middle column of Figure~\ref{fig:outprops} shows the outflow momentum flux as a function of SFR and \LAGN. We find only modest momentum `boosts' above the radiative momentum flux provided by star formation or the AGN individually, and a median $\pdot/(\lir/c) = 0.9$ when taken together. The molecular outflows in the $z>6$ quasars we have observed are thus fully consistent with being momentum-driven winds, regardless of their ultimate driving source(s). This is in contrast to some of the most extreme molecular outflows observed at $z\sim0$, which can show momentum boosts up to $\sim10$, likely evidence of an energy-conserving wind phase \citep[e.g.][]{gonzalezalfonso17}. Alternatively if the quasar outflows are radiation-driven, high dust opacities (i.e. high gas and dust column densities) are not necessary to explain the momentum of the cold outflow phase.

Based on the outflow momentum flux alone, we cannot distinguish whether the outflows are driven by star formation or the AGN; the observed range is compatible with either scenario. The kinetic power in the molecular outflows can help distinguish which of these is ultimately most plausibly responsible for driving the outflows. Assuming that the supernova rate scales linearly with the SFR and a Kroupa IMF, supernovae can produce $\dot{E}_{\mathrm{SNe}} \approx 1.1\times10^8 \times \mathrm{SFR}$ \Lsol, with SFR in \Msol/yr \citep{leitherer99,veilleux05}. Only a small fraction $\lesssim$10\% of the total supernova energy is expected to couple directly into kinetic motion of the outflowing gas \citep[e.g.][]{kim15,gentry17}. In AGN-driven winds, up to a maximum of 2--5\% of the total AGN luminosity is thought to be able to couple directly to kinetic motion of the outflowing gas across all gas phases \citep[e.g.][]{king15}.

The right panels of Figure~\ref{fig:outprops} show the outflow kinetic power, again as a function of SFR and \LAGN. Compared to the power available from supernovae, the outflows span two orders of magnitude from $\approx1-100$\%. On average, the outflows would require $\approx$30\% of the supernova power to couple to the cold molecular phase alone; warmer outflow phases would increase this requirement even further. In contrast, even the most extreme $z>6$ quasar outflow requires at most $\approx$1\% of the AGN power to couple to the cold gas; 0.1\% \LAGN is more typical. This is well below theoretical expectations for the maximum coupling efficiency. These fractions would increase somewhat if we used the IR-based AGN luminosities, which are typically $\approx2\times$ lower, but the outflow energetics would still be well below the theoretical maximum.

Are the molecular outflows driven by radiation pressure on dust, or hydrodynamical shock forces? Given the large uncertainties in the observations and flexibility in the models, this is hard for us to distinguish. Perhaps our best evidence against radiation pressure is that the outflow energetics do not seem to follow the predicted $\Edot \propto L^{3/2}$ scaling \citep{ishibashi18}. This remains true if we restrict the low-redshift samples to include only Seyferts, or if we require that the AGN be the dominant energy source in the system ($\fagn > 0.5$), or if we used the IR-based AGN luminosities for our quasar sample. In all cases, the outflow kinetic power is approximately linear or sub-linear with \LAGN. There is, however, substantial flexibility in the models to accommodate a wide range of observations. The followup models of \citet{ishibashi22} can span more than two orders of magnitude in $\Edot/\LAGN$ by appealing to AGN variability and/or underestimated AGN luminosities due to high obscuration, for example. We thus cannot rule out radiation pressure as the outflow driving mechanism.

Taken together, the right panels of Figure~\ref{fig:outprops} provide compelling evidence for AGN-driven feedback on the cold molecular gas in $z>6$ quasars. Supernovae alone cannot power the outflows unless an unrealistically high fraction $\gtrsim$30\% of all the energy they release couples directly into motion of the cold gas. On the other hand, there is easily enough energy supplied by the AGN to power the observed molecular outflows. This result is primarily due to a direct observable, the outflow velocity \vout. The faster outflow velocities compared to the DSFGs result in more powerful outflows even if the outflow rates are basically the same. With the present data, we have shown that the AGN are likely responsible for driving the outflows, and that energy-conserving wind phases are not required to explain the energetics.

\subsection{Cold outflows and black hole accretion} \label{mbh}

We have argued that at least some energy injection from the AGN is needed to power the molecular outflows we have observed. Given this, it is interesting to look for trends between cold outflow properties and the properties of the central supermassive black holes. For the $z>6$ quasars and the low-redshift OH and CO outflow samples, we compiled estimates of the supermassive black hole mass. For the high-z quasars, \MBH is derived from single-epoch virial estimates based on the MgII and/or CIV broad lines following standard scaling relations \citep{andika20,yang21,farina22}. For the low-redshift samples several methods were used. For a small number of very nearby galaxies, \MBH was derived from the kinematics of the nuclear regions ($r<10$\,pc). The large majority, however, come from applying variants of the $M-\sigma$ relation, either using the central velocity dispersion or the central K-band luminosity \citep[e.g.][]{veilleux09b}. The uncertainties on individual measurements are obviously considerable; we adopt 0.5\,dex uncertainties in the following.

\begin{figure}
\includegraphics[width=\columnwidth]{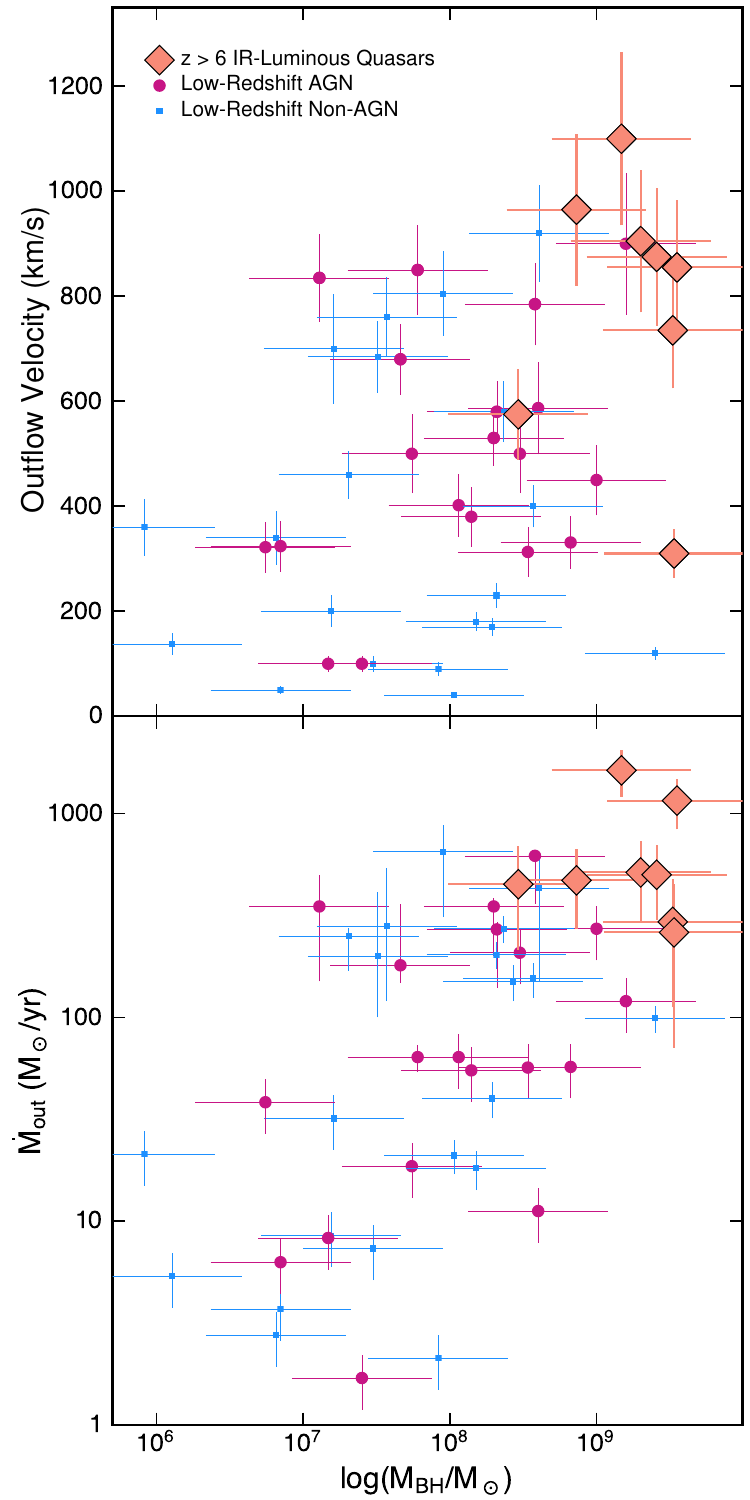}
\caption{
Molecular outflow velocity (top) and outflow rate (bottom) as a function of supermassive black hole mass \MBH. We now distinguish the low-redshift comparison samples classified as either Type~1 or 2 AGN from the non-AGN galaxies. We see weak trends with \MBH, but this is most likely due to correlations between \MBH and other galaxy properties.
}\label{fig:mbh}
\end{figure}

Figure~\ref{fig:mbh} shows the molecular outflow velocities and outflow rates as a function of \MBH for the low- and high-redshift samples.  While we do not expect this to be a particularly meaningful comparison, we include it here for completeness. Unlike in previous figures, we now distinguish the low-redshift sample galaxies classified as Type 1 or 2 AGN vs. otherwise, rather than by the provenance of the molecular outflow estimate. Our goal is to isolate galaxies in which the AGN likely plays a prominent role in the overall energetics.\footnote{We find qualitatively similar results if we divide the sample based on \fagn.}

Figure~\ref{fig:mbh} shows, as expected, considerable scatter in both outflow velocity and rate at a given \MBH, similar to past low-redshift studies \citep[e.g.][]{rupke17}. We find no correlation in either panel within the $z>6$ quasars alone; these quasars span a very small dynamic range in \MBH. We do see a weak trend of higher \Mdot at higher \MBH, but this is plausibly due to the fact that more massive galaxies host more massive black holes, higher SFRs, and (at least in this sample) more luminous AGN. In many ways, this reflects a common outcome in extragalactic astronomy: bright galaxies are bright; in this case galaxies bright enough to have black hole mass estimates (from a variety of methods) are also bright enough to have outflow rate estimates.

\begin{figure}
\includegraphics[width=\columnwidth]{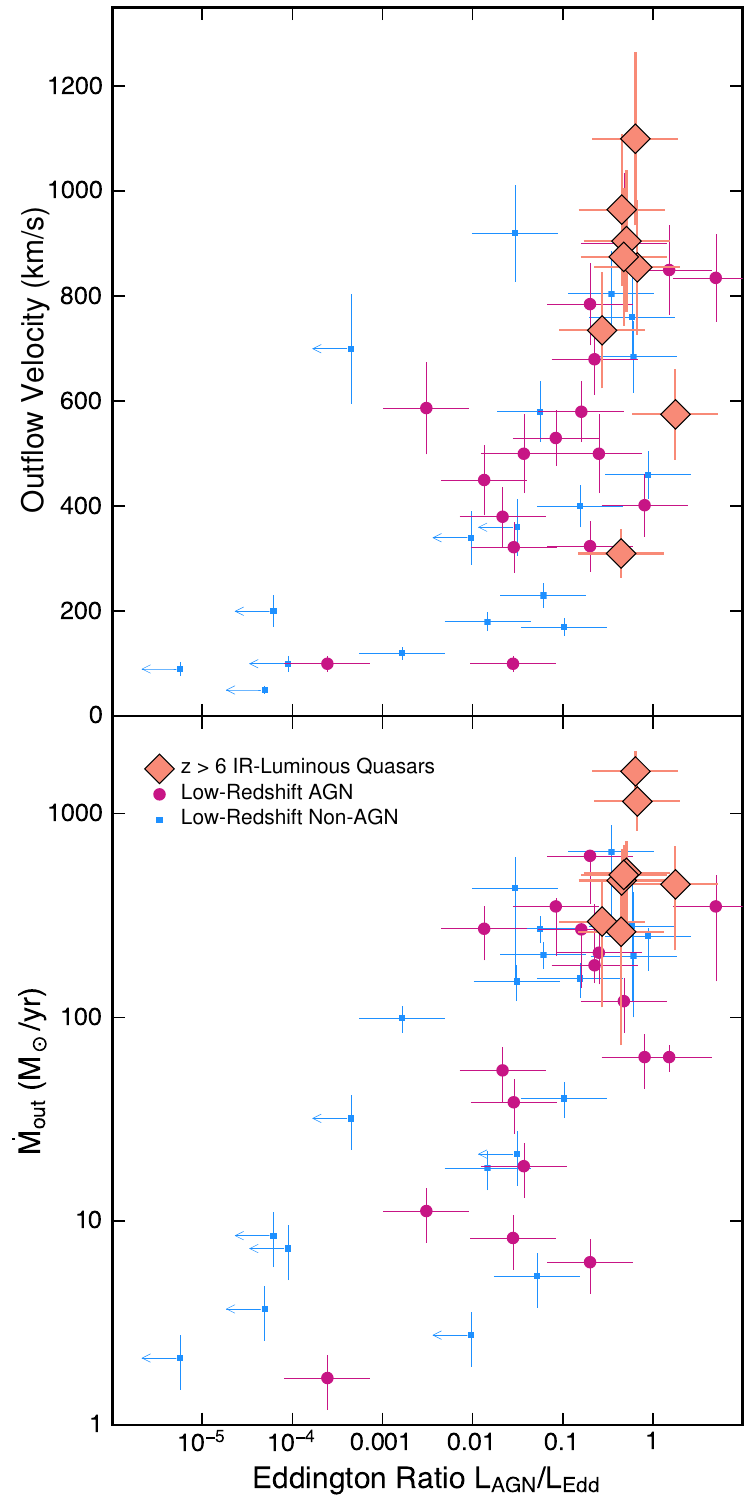}
\caption{
Outflow properties as a function of Eddington ratio, with symbols as in Fig.~\ref{fig:mbh}. Because the x-axis incorporates \LAGN, which we have already shown is strongly correlated with the outflow properties, we note strong trends in both panels. 
}\label{fig:lamedd}
\end{figure}

Figure~\ref{fig:lamedd} shows the same outflow quantities as a function of the Eddington ratio \LAGN/$L_{\mathrm{Edd}}$. Here we see much stronger trends with both \vout and \Mdot. These correlations were entirely expected, because we have already shown that both \vout and \Mdot are correlated with \LAGN itself (Figs.~\ref{fig:voutlagn} and \ref{fig:outprops}). 

It is interesting to note that the scatter in Fig.~\ref{fig:lamedd} appears larger in the low-redshift galaxies that are \textit{not} classified as AGN (i.e. the blue points). The obvious conclusion is that their outflows are more likely driven by star formation. Considering only the AGN, \Mdot and Eddington ratio both increase proportionally by three orders of magnitude over the observed range. Though it is difficult to think of how such sources might be selected in large numbers, it is clear that high-redshift quasars that span a much wider range in Eddington ratio will be needed to understand if and how these quantities relate to the coldest phase of galactic outflows.

The black hole mass \MBH is, in some sense, a measure of the cumulative energy released by the AGN over the history of each galaxy's formation, and thus averages over variability in the AGN luminosity. The typical outflow dynamical time $\approx1-10$\,Myr is much longer than the typical timescale for AGN accretion $\sim0.1$\,Myr \citep[e.g.][]{schawinski15}, so the observed AGN-driven winds likely average over many past accretion episodes. This is especially relevant for the low-redshift samples, where \MBH integrates the AGN accretion over 13\,Gyr of cosmic history. It will be interesting to learn whether high-redshift quasars, with sufficient dynamic range in \MBH, show a tighter correlation with \MBH due to the shorter history of the universe at such early times.

\subsection{Comparison to other potential outflow tracers} \label{outcomparison}

Our OH observations trace the cold and dense phase of galactic winds, but these outflows are expected to contain gas over many orders of magnitude in temperature and density \citep[e.g.][]{schneider18}. If the observed quasar outflows are ultimately driven by thermal energy or radiation pressure from the black hole accretion disks, one might expect trends between the large-scale cold outflows and warm or hot wind phases arising from closer to the central engines. In this section we compare the cold outflow phase to other outflow tracers. Throughout this work we have endeavored to treat our $z>6$ quasar sample as homogeneously as possible. Due to the varying availability and quality of the ancillary multiwavelength data, we cannot do the same here, and a detailed investigation of multiphase wind tracers is beyond the scope of this work. Instead we give only brief overviews of different outflow tracers and the observations that are currently available for our sample.

\begin{figure}
\includegraphics[width=\columnwidth]{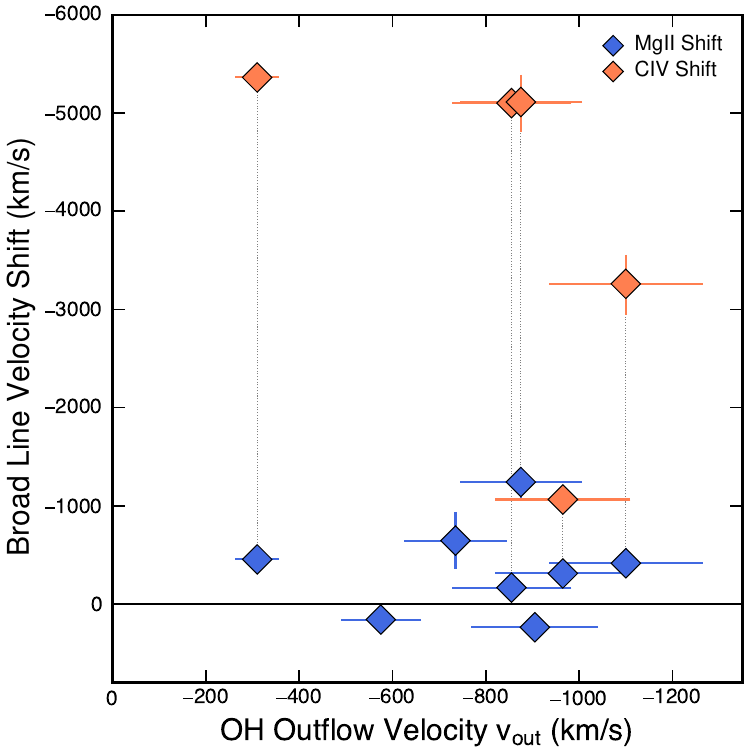}
\caption{
We find no correlation between the OH molecular outflow velocity and the velocity shifts in the MgII (blue) or CIV (orange) broad lines observed in the near-IR. If the broad-line shifts arise due to a hot nuclear wind component, this component is not strongly coupled to the kpc-scale cold outflow. All velocity shifts are relative to the ALMA \cii redshift. Dotted gray lines connect quasars for which both velocity MgII and CIV velocity shifts were measured.
}\label{fig:voutblr}
\end{figure}

\paragraph{Broad Line Region Tracers}
Velocity shifts of quasar broad emission lines have been known for decades and have inspired an entire literature \citep[e.g.][]{gaskell82,vandenberk01,richards11}. Broad, blueshifted CIV, in particular, might arise from a wind launched from the accretion disk \citep[e.g.][]{murray95}, but velocity shifts are also observed in low-ionization tracers like MgII. If the velocity shifts are caused by hot nuclear winds and the large-scale molecular outflows are driven by those winds, we might find correlations between the broad-line velocity shifts and the cold outflow properties. Figure~\ref{fig:voutblr} compares the blueshifted OH outflow velocity to the velocity shifts observed in the rest-UV MgII and CIV lines for our quasars. The broad-line shifts are assembled from the literature \citep{schindler20,andika20,yang21}; all velocity shifts are relative to the ALMA \cii redshift, which we take to trace the systemic velocity. We see no evidence of a correlation between either broad-line tracer and the OH blueshift. If the broad line shifts do arise from nuclear winds, they are not closely connected to the cold outflow on large scales; alternatively, it may be that the broad line shifts do not arise from fast nuclear outflows.

\paragraph{Warm Ionized Outflows} 
Similarly, the warm ionized phase of quasar outflows is traced by broad rest-optical lines that emerge from larger galactic scales, especially \oiii. At lower redshifts $z\sim1-2$, \oiii remains in the near-IR atmospheric windows observable from the ground, and ionized outflows have been detected in large numbers of quasars this way \citep[e.g.][]{nesvadba08,perna15}. This redshift range is particularly valuable because the cold molecular phase of outflows can be traced with ALMA observations of CO \citep{vayner21a}. Higher-redshift extensions have awaited the arrival of JWST's longer-wavelength spectroscopic modes. At present, 5/11 quasars in our sample have already been observed with JWST spectroscopy. Of these, the rest-optical spectra have been published by \citet{yang23} for two, J0305-3150 and P036+03 (aka J0226+0302). Neither source shows evidence for \oiii outflows, but neither source has an OH outflow either. Data for the other 3 quasars are not yet published. The existing and upcoming JWST spectroscopy should allow more comprehensive comparisons between the cold and warm ionized outflow phases.

\paragraph{Cool Neutral Atomic Outflows}
More closely related to the cold molecular outflows we have detected, \cii might also trace the cool/cold phase of galactic outflows via broad line wings. Unfortunately \cii is a rather controversial outflow tracer. The wings, if present, are faint, motivating stacking analyses of many galaxies to bring out the weak signal. In $z>6$ quasars in particular, past stacking analyses have come to opposite conclusions about the presence or absence of broad \cii wings even though many of the input quasars were the same \citep{decarli18,bischetti19,stanley19,novak20}. The same is true for individual quasars, with at least some well-known past claims of broad \cii wings \citep{maiolino12,cicone15} vanishing under more scrutiny \citep{meyer22}. Other galaxies with clear OH outflows show no evidence of broad \cii line wings \citepalias{spilker20a}. Careful methods and high-quality data are clearly required for robust conclusions. 

Although all quasars in our sample were selected to have existing ALMA \cii detections, the quality of the \cii data is highly variable in depth and resolution, and some quasars have been observed in \cii several times. It is beyond the scope of this work to combine and analyze all the \cii data available for each source.  It is clear from Fig.~\ref{fig:ohspectra}, however, that the outflows are far more pronounced in OH than \cii. A couple quasars show signs of broad \cii wings extending to $\approx$500\,\kms (e.g. P183+05, P231-20). The most prominent OH outflows in our sample, with $\vout>1000$\,\kms (J0439+1634, J2310+1855), show little evidence of \cii emission at such high velocities. The physical reasons that \cii does not trace molecular outflows reliably require further investigation.

\subsection{OH 119\,\um in emission} \label{ohemission}

We have almost exclusively focused on the blueshifted OH 119\,\um absorption features in the quasar spectra. Most quasars, however, show this line in emission either alone (2/8 quasars) or together with absorption features (4/8). Including the three literature objects, $\approx82\pm11$\% of $z>6$ quasars show OH 119\,\um in emission.\footnote{This includes J1319+0950 \citepalias{herreracamus20}, who invoke OH emission `canceling out' the doublet of the OH absorption.} This is very different than the $z\sim5$ DSFG sample, in which 0/11 galaxies showed OH emission ($<$27\% at 95\% confidence level from Poisson statistics). The emission features we detect are not particularly broad (Fig.~\ref{fig:ohspectra}), and so are probably not associated with fast gas flows.

Producing OH 119\,\um in emission requires very warm and dense gas to populate the upper energy levels due to the short radiative lifetime (large Einstein `A' coefficient) of the upper levels. While in principle OH 119\,\um could be in emission due to either radiative pumping of the upper energy levels or collisional excitation, the observed ratios of different OH transitions at low redshift suggest that collisional excitation dominates \citep{spinoglio05,runco20}. 

The warm and dense gas that produces OH emission may be found near the nuclear regions of quasar host galaxies. Indeed it is clear from large low-redshift samples that OH 119\,\um emission is associated with AGN-dominated systems, including obscured AGN \citep{veilleux13,stone16,runco20}. The presence of OH emission in the $z>6$ quasars and its absence in the $z\sim5$ DSFGs is therefore consistent with the relative dominance of the AGN. 

Although OH 119\,\um emission is clearly correlated with the presence of an AGN within the galaxy, it does not appear to arise solely from within the nuclear region itself. Due to the low spatial resolution of Herschel/PACS, low-redshift arguments in favor of OH emission on wider $\sim100-200$\,pc scales center around radiative transfer models and correlations with the silicate dust opacity feature \citep[e.g.][]{spinoglio05,stone16,gonzalezalfonso17}. 

At high redshift, the superior spatial resolution of ALMA can directly resolve the scale of the OH emission region. \citet{salak24} used $\approx$0.2\arc observations to estimate a $\approx$2\,kpc OH-emitting region in the quasar J2054-0005. For the four quasars in our sample with the highest S/N, the OH emission is marginally resolved (not pointlike) in three of them (J0305-3150, J0525-2406, P183+05; P009-10 is pointlike at our data quality). The peak of the emission is coincident with the continuum peak in all cases, within the uncertainties. The beam-deconvolved sizes (circularized FWHM) of the emitting regions range from 2--8\,kpc with large uncertainties, estimated by fitting a 2D Gaussian profile to the images in Fig.~\ref{fig:images}. 

Though somewhat larger than the size derived by \citet{salak24}, our results are in qualitative agreement that the OH emission is not produced solely within the nuclear region. The larger sizes we find are most likely due to the poor spatial resolution of our data and differences in fitting methodology. The fact that OH 119\,\um emission is clearly associated with AGN but arises on larger galactic scales implies that either AGN are simply more common in galaxies with sufficiently warm and dense gas, or that the AGN impacts the host-scale ISM to produce the warm/dense conditions needed for OH 119\,\um to be in emission.

\section{Conclusions} \label{conclusions}

We have assembled the first modest-sized sample of $z>6$ quasars with observations of the molecular outflow tracer OH 119\,\um, combining 8 new quasars with 3 from the literature. The sample is not representative of all known $z>6$ quasars; we targeted the far-IR brightest objects as an initial foray. OH is clearly detected in all targets in either emission, absorption, or both. This is presently the largest sample of high-redshift quasars with molecular outflow constraints. Our main results include:

\begin{itemize}

\item We detect unambiguous molecular outflows in 73\%\,$\pm$\,13\% of $z>6$ IR-luminous quasars. Because this is a lower limit to the true incidence rate of outflows, it is clear that cold outflows are ubiquitous among the most IR-luminous high-redshift quasars. 

\item The molecular outflows are substantially faster in $z>6$ quasars than $z\sim5$ DSFGs at comparable luminosity, by over 300\,\kms on average. We find a clear correlation between the outflow velocity \vout and AGN luminosity \LAGN when combined with low-redshift comparison samples. We take this as direct evidence for AGN feedback on the cold gas in $z>6$ quasars.

\item From optical depth arguments, we conclude that the quasar outflows are more compact relative to their host galaxies than either the $z\sim5$ DSFG or low-redshift samples. This could be because they have smaller opening angles, or fewer/smaller clumps of outflowing cold gas. Detailed constraints on the structure of the outflows await higher-resolution ALMA campaigns.

\item The $z>6$ quasar outflows are not otherwise unusually strong; in fact the OH absorption is weaker than in similar-luminosity DSFGs. The consequence is that the molecular outflow rates \Mdot are rather modest given the extreme luminosities, with wind mass-loading factors $\etaout = \Mdot/\mathrm{SFR} \approx 0.5$. 

\item The physical properties of the molecular outflows are highly uncertain given the available data, but they are consistent with expectations for momentum-driven winds powered either by star formation or the AGN. The outflow kinetic power, however, would require an unrealistically high fraction of the power produced by supernovae to be converted into bulk outflow motion; in contrast the AGN supply more than sufficient power to drive the observed outflows based on theoretical expectations. This result is a direct consequence of the faster outflow speeds we observe in  the quasars compared to literature samples.

\item We identify trends between the outflow properties, the supermassive black hole mass \MBH, and the SMBH Eddington ratio, in agreement with our finding that the AGN are at least partly responsible for driving the outflows. We stress, however, that these trends may be driven by underlying confounding quantities such as \LAGN.

\end{itemize}

Our findings strongly suggest that the AGN play a large role in driving the cold molecular phase of outflows in high-redshift quasars, though we are far from `smoking gun' evidence. This conclusion is primarily based on a direct observable quantity -- the faster blueshifted velocities in comparison to high-redshift non-quasars. Given the high detection rate of unambiguous molecular outflows, extending our work to lower-luminosity systems is a promising path for future studies, because even modest sample sizes are likely to find enough outflows to place meaningful statistical constraints on their properties.

\begin{acknowledgements}
We thank the referee for their careful reading of the paper, which helped to shore up many of the arguments presented here.

This paper makes use of the following ALMA data: 2016.1.01063.S, 2018.1.01289.S, 2019.1.01025.S, 2019.2.00053.S, 2021.1.00443.S, 2021.2.00064.S, 2021.2.00151.S. 
ALMA is a partnership of ESO (representing its member states), NSF (USA) and NINS (Japan), together with NRC (Canada), MOST and ASIAA (Taiwan), and KASI (Republic of Korea), in cooperation with the Republic of Chile. The Joint ALMA Observatory is operated by ESO, AUI/NRAO and NAOJ. The National Radio Astronomy Observatory is a facility of the National Science Foundation operated under cooperative agreement by Associated Universities, Inc.

This work was performed in part at the Aspen Center for Physics, which is supported by National Science Foundation grant PHY-1607611.

This research has made use of NASA's Astrophysics Data System.
\end{acknowledgements}

\facility{ALMA}

\software{
CASA \citep{mcmullin07},
\texttt{astropy} \citep{astropy18},
\texttt{matplotlib} \citep{hunter07}}

\clearpage
\bibliographystyle{aasjournal}

\restartappendixnumbering
\appendix

\section{Far-Infrared Photometry and SED Fits} \label{appsedfits}

As described briefly in Section~\ref{ancillary}, we compiled the available far-IR photometry for the quasars in our sample in order to measure \lir and \lfir more robustly than previous estimates. Some quasars in our sample have been extensively studied at many wavelengths, while others have few additional measurements. At minimum, all quasars are strongly detected at rest-frame 160 and 120\,\um continuum, because we selected the quasars to have bright 160\,\um continuum from existing \cii observations and add the 120\,\um from our own data. When possible, we use published continuum measurements taken from the literature. In cases where additional ALMA data are publicly available from the archive but as yet unpublished, we downloaded the pipeline full-bandwidth continuum images and measured flux densities using the CASA \texttt{imfit} task just as we do for our OH data. For all quasars but J0525-2406, there is at least one measurement at higher frequency than our rest-120\,\um data, which allows for better constraints near the peak of the dust SED. 

Table~\ref{tab:firphot} provides the continuum photometry we assembled from the literature and publicly-available archival data. Note that the flux densities from our ALMA data in this table differ from those given in Table~\ref{tab:ohfits}. In Table~\ref{tab:ohfits} we estimated the continuum flux density directly at the wavelength of the upper OH transition, while here the continuum combines both sidebands of the ALMA data.

\begin{figure*}[t]
\begin{centering}
\includegraphics[width=0.24\textwidth]{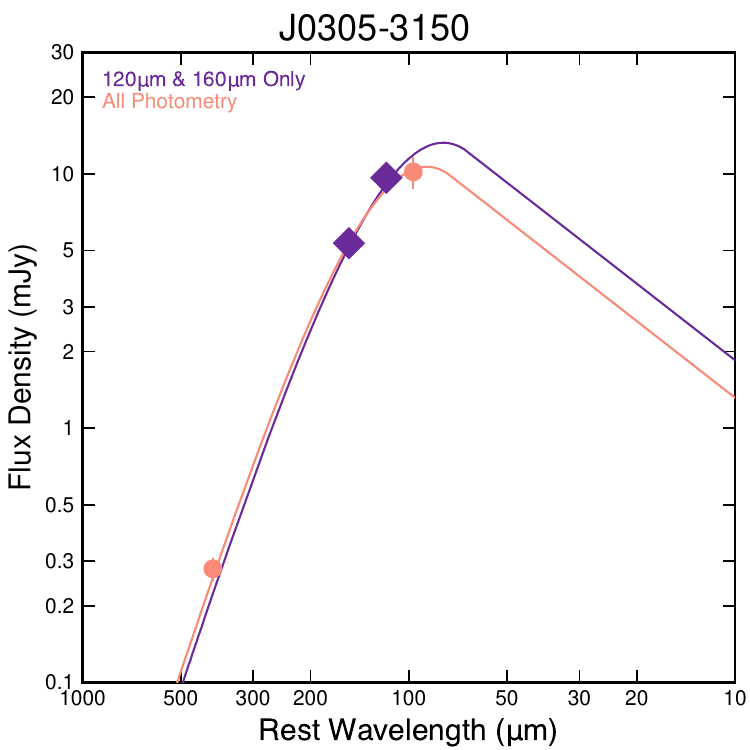}
\includegraphics[width=0.24\textwidth]{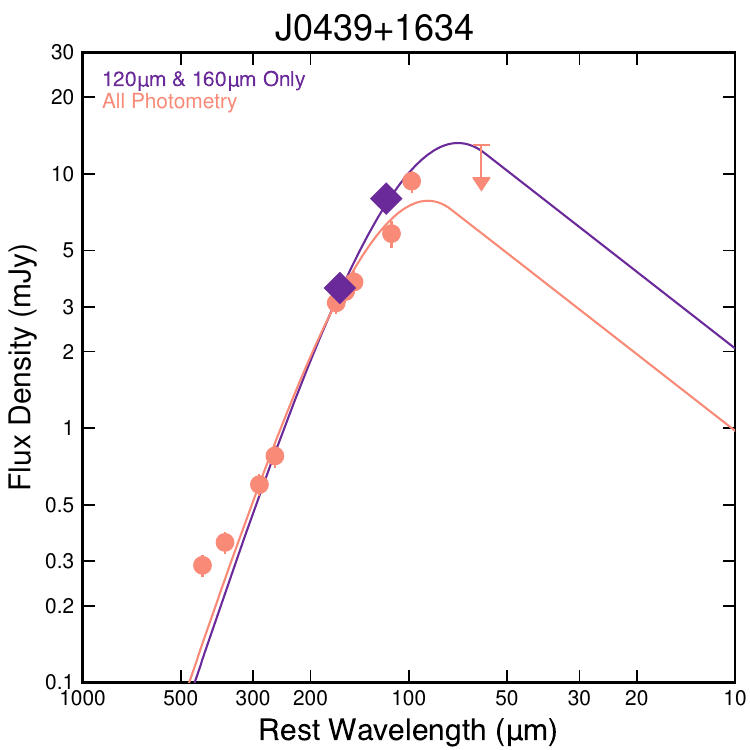}
\includegraphics[width=0.24\textwidth]{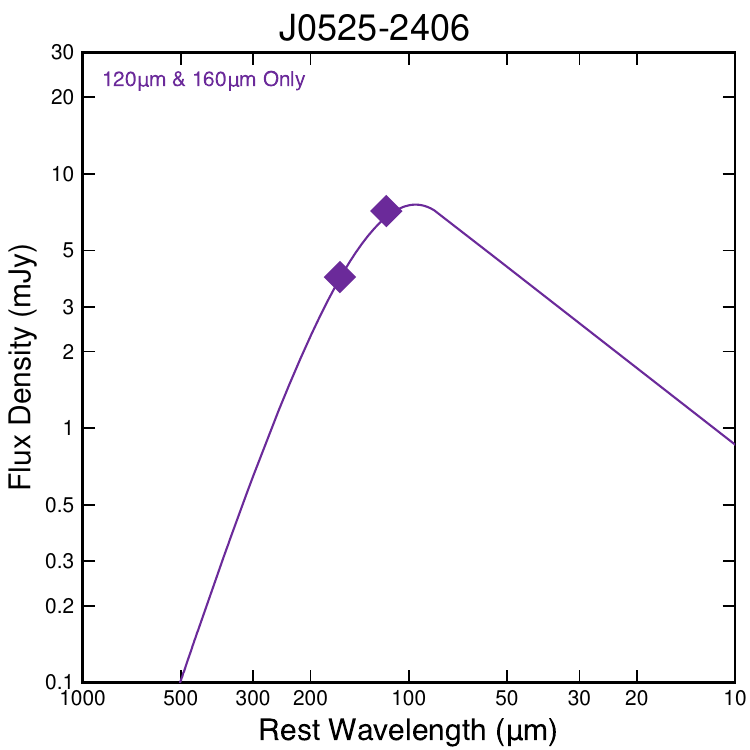}
\includegraphics[width=0.24\textwidth]{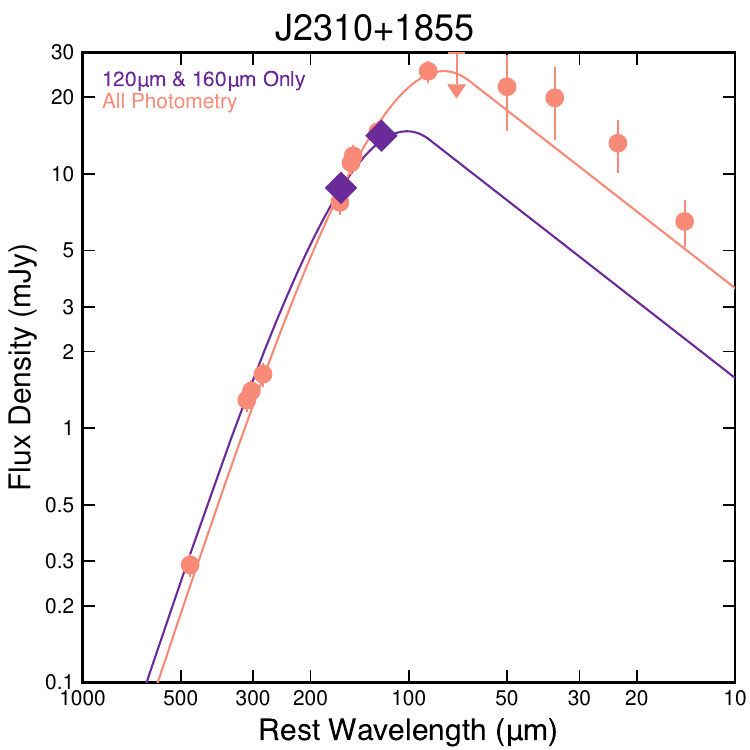}
\includegraphics[width=0.24\textwidth]{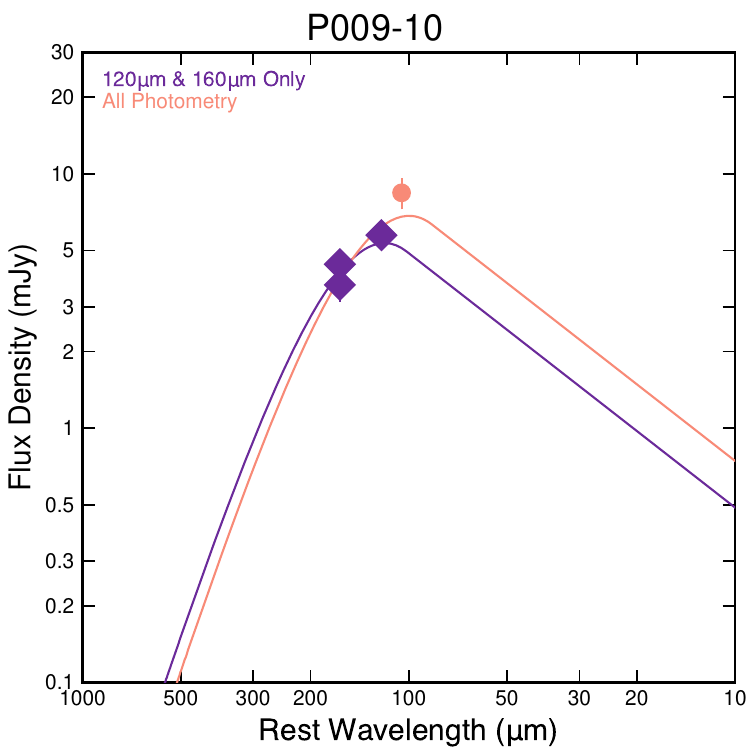}
\includegraphics[width=0.24\textwidth]{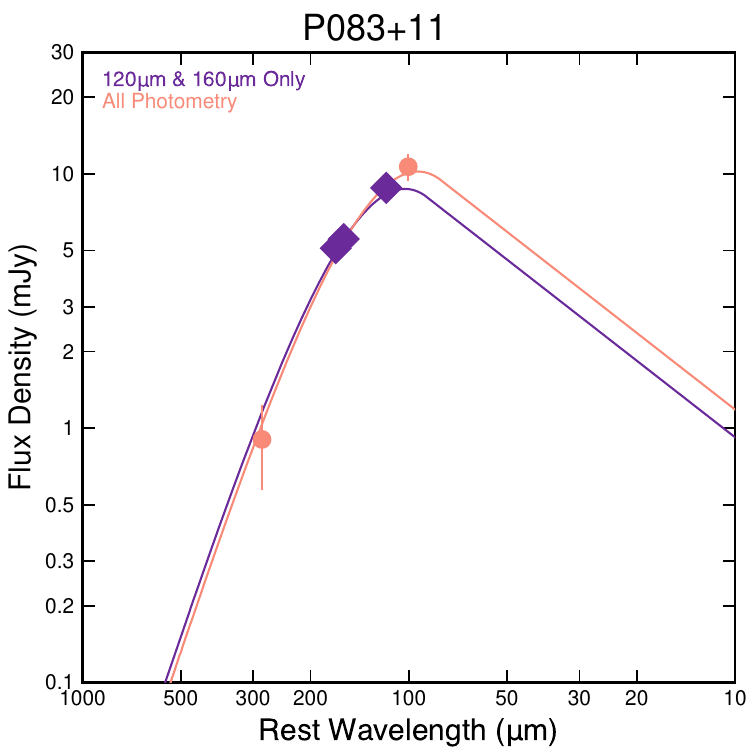}
\includegraphics[width=0.24\textwidth]{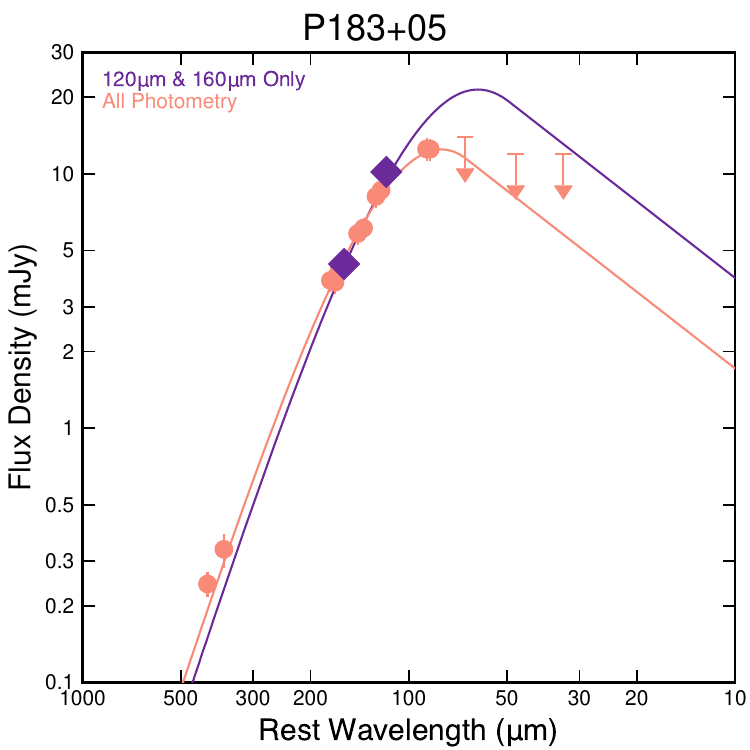}
\includegraphics[width=0.24\textwidth]{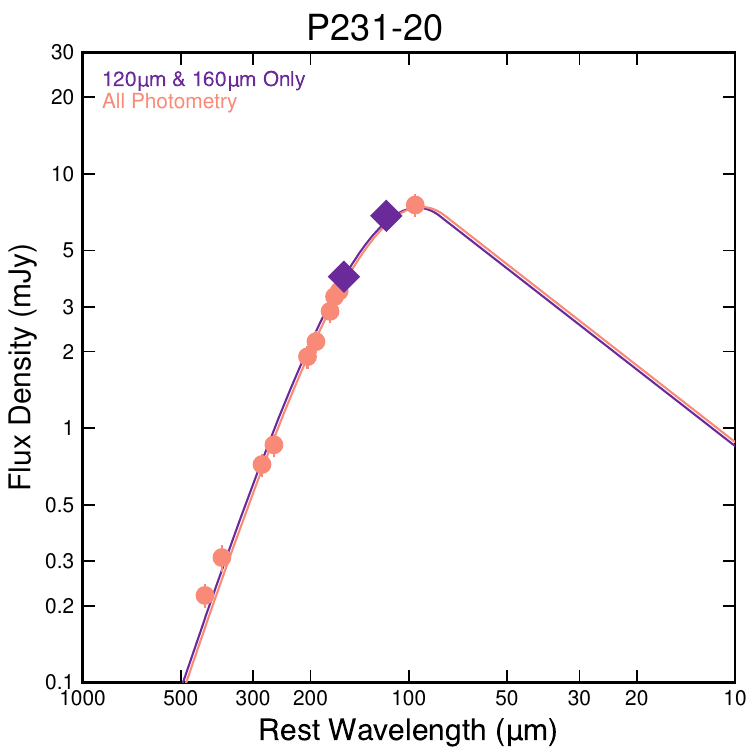}
\includegraphics[width=0.24\textwidth]{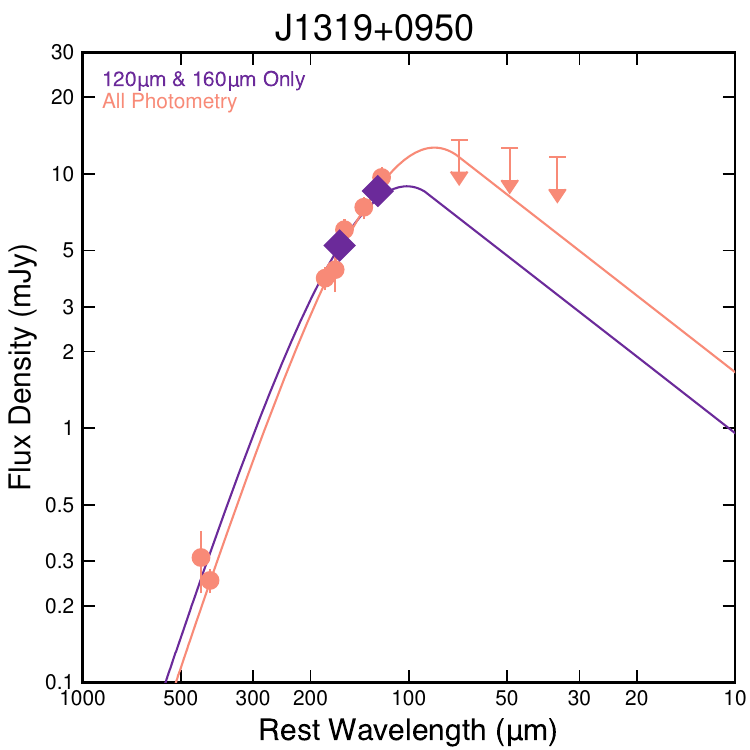}
\includegraphics[width=0.24\textwidth]{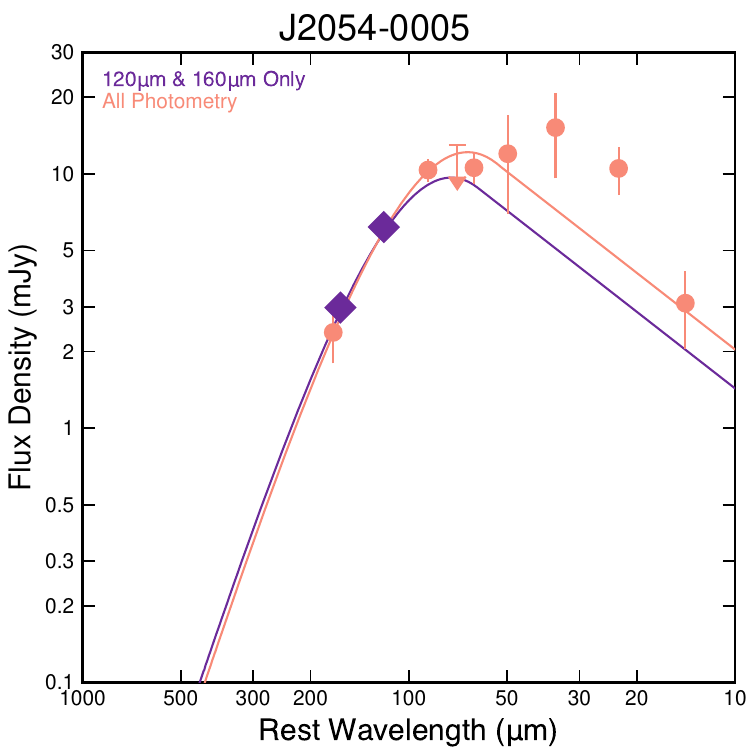}
\includegraphics[width=0.24\textwidth]{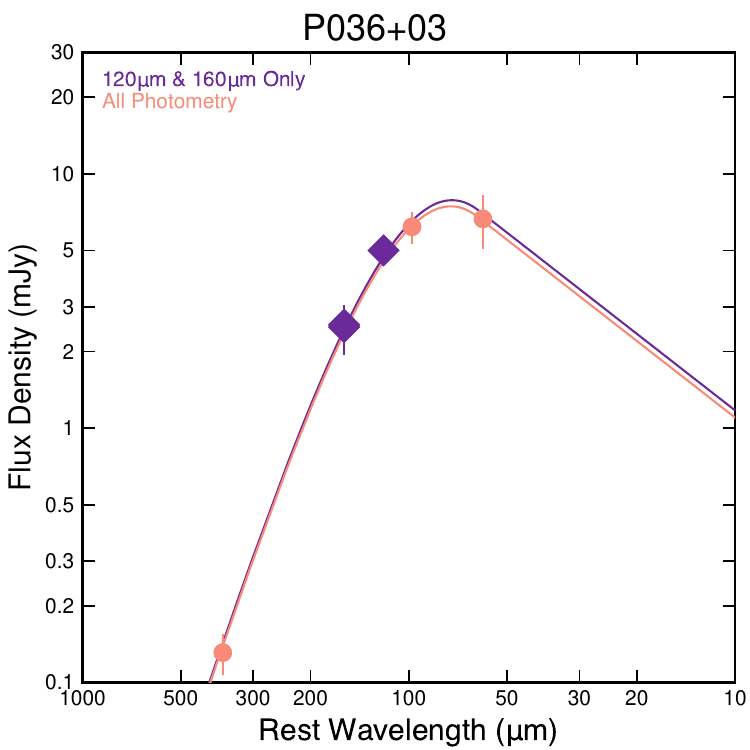}\\
\end{centering}
\caption{
Fits to the far-IR SEDs of our sample quasars. Navy points and lines show the rest-frame 120 and 160\,\um ALMA measurements available for all quasars and a fit to only these points. Peach points and lines include all additional far-IR continuum measurements from the literature and/or the ALMA archive. Upper limits are at the 2$\sigma$ level.
}\label{fig:appsedfits}
\end{figure*}

\begin{figure}[h]
\begin{centering}
\includegraphics[width=0.4\textwidth]{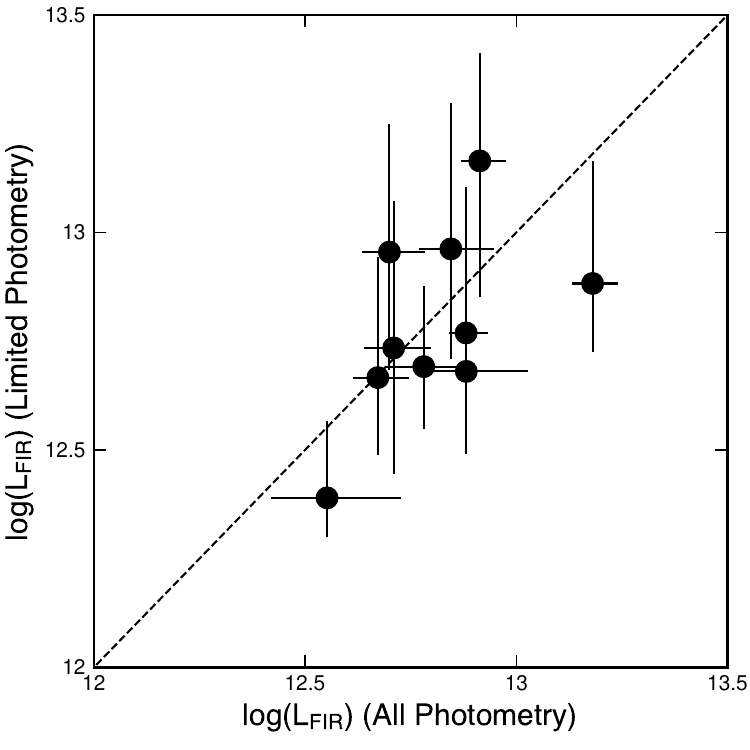}\\
\end{centering}
\caption{
The luminosity estimates for our quasar sample are consistent within the uncertainties between the two sets of SED fits we perform. We find no evidence for systematic offsets or biases when the additional far-IR photometry is included or excluded.
}\label{fig:applfir}
\end{figure}

Figure~\ref{fig:appsedfits} shows the best fit SEDs using the model described in Sec.~\ref{sedfitting}, including one fit to only the 120 and 160\,\um measurements available for all sources and one fit to all available data. Figure~\ref{fig:applfir} compares the measured luminosities between these fits; J0525-2406 is not present because it has no additional far-IR measurements.

\startlongtable
\begin{deluxetable}{lDDc}
\tablecaption{Compilation of Long-wavelength Photometry \label{tab:firphot}}
\tablehead{
\colhead{Source} &
\multicolumn2c{$\nu_{\mathrm{obs}}$ (GHz)} &
\multicolumn2c{$S_\nu$ (mJy)} &
\colhead{Reference}} 
\decimals
\startdata
J0305-3150 &   98.8 &  0.28 $\:\pm$ 0.03 &   (1)          \\
           &  258.1 &  5.20 $\:\pm$ 0.53 &   (1)          \\
           &  336.1 &  9.65 $\:\pm$ 0.97 & 2021.1.00443.S \\ 
           &  406.8 & 10.18 $\:\pm$ 1.42 & 2019.2.00053.S \\ 
J0439+1634 &   93.0 &  0.29 $\:\pm$ 0.03 &   (2)          \\
           &  109.0 &  0.36 $\:\pm$ 0.04 &   (2)          \\
           &  139.0 &  0.60 $\:\pm$ 0.06 &   (2)          \\
           &  155.0 &  0.78 $\:\pm$ 0.08 &   (2)          \\
           &  239.0 &  3.11 $\:\pm$ 0.31 &   (2)          \\
           &  245.2 &  3.56 $\:\pm$ 0.36 &   (3)          \\
           &  255.0 &  3.44 $\:\pm$ 0.35 &   (2)          \\
           &  271.0 &  3.76 $\:\pm$ 0.37 &   (2)          \\
           &  340.3 &  7.99 $\:\pm$ 0.80 & 2021.1.00443.S \\ 
           &  353.0 &  5.82 $\:\pm$ 0.70 &   (2)          \\
           &  407.0 &  9.36 $\:\pm$ 0.94 & 2021.2.00064.S \\ 
           &  666   &       $<$6.47      &   (2)          \\
J0525-2406 &  244.7 &  3.93 $\:\pm$ 0.40 & 2019.1.01025.S \\ 
           &  339.4 &  7.14 $\:\pm$ 0.72 & 2021.1.00443.S \\ 
\tablebreak
J2310+1855 &   91.5 &  0.29 $\:\pm$ 0.03 &   (4)          \\
           &  136.6 &  1.29 $\:\pm$ 0.13 &   (4)          \\
           &  141.0 &  1.40 $\:\pm$ 0.14 &   (4)          \\
           &  153.1 &  1.63 $\:\pm$ 0.17 &   (4)          \\
           &  263.3 &  7.73 $\:\pm$ 0.83 &   (4)          \\
           &  265.4 &  8.81 $\:\pm$ 0.89 &   (4)          \\
           &  285.0 & 11.05 $\:\pm$ 1.12 &   (4)          \\
           &  289.2 & 11.77 $\:\pm$ 1.18 &   (4)          \\
           &  344.2 & 14.63 $\:\pm$ 1.50 &   (4)          \\
           &  353.1 & 14.12 $\:\pm$ 1.41 & 2021.1.00443.S \\ 
           &  490.8 & 25.31 $\:\pm$ 2.54 &   (4)          \\
           &  600   &       $<$15.0      &   (4)          \\
           &  857   & 22.0  $\:\pm$ 7.2  &   (4)          \\
           & 1200   & 19.9  $\:\pm$ 6.3  &   (4)          \\
           & 1873   & 13.2  $\:\pm$ 3.1  &   (4)          \\
           & 3000   &  6.5  $\:\pm$ 1.4  &   (4)          \\
P009-10    &  263.3 &  4.41 $\:\pm$ 0.50 &   (5)          \\
           &  263.3 &  3.66 $\:\pm$ 0.51 &   (6)          \\
           &  353.0 &  5.74 $\:\pm$ 0.57 & 2021.1.00443.S \\ 
           &  406.9 &  8.42 $\:\pm$ 1.18 & 2019.2.00053.S \\ 
P083+11    &  145.0 &  0.90 $\:\pm$ 0.33 & 2021.2.00151.S \\ 
           &  244.0 &  5.10 $\:\pm$ 0.53 &   (7)          \\
           &  258.0 &  5.54 $\:\pm$ 0.58 &   (7)          \\
           &  348.5 &  8.81 $\:\pm$ 0.88 & 2021.1.00443.S \\ 
           &  407.0 & 10.67 $\:\pm$ 1.28 & 2021.2.00064.S \\ 
P183+05    &   97.4 &  0.24 $\:\pm$ 0.03 &   (8)          \\
           &  109.3 &  0.33 $\:\pm$ 0.05 &   (8)          \\
           &  231.9 &  3.81 $\:\pm$ 0.38 &   (8)          \\
           &  239.5 &  3.75 $\:\pm$ 0.38 &   (8)          \\
           &  246.6 &  4.33 $\:\pm$ 0.44 &   (8)          \\
           &  255.5 &  4.42 $\:\pm$ 0.45 &   (8)          \\
           &  280.9 &  5.82 $\:\pm$ 0.58 &   (8)          \\
           &  293.0 &  6.12 $\:\pm$ 0.61 &   (8)          \\
           &  319.8 &  8.15 $\:\pm$ 0.82 &   (8)          \\
           &  331.7 &  8.63 $\:\pm$ 0.86 &   (8)          \\
           &  343.9 & 10.17 $\:\pm$ 1.02 & 2021.1.00443.S \\ 
           &  457.0 & 12.52 $\:\pm$ 1.26 &   (8)          \\
           &  469.1 & 12.51 $\:\pm$ 1.26 &   (8)          \\
           &  600   &       $<$7.0       &   (9)          \\
           &  857   &       $<$6.0       &   (9)          \\
           & 1200   &       $<$6.0       &   (9)          \\
P231-20    &   93.8 &  0.22 $\:\pm$ 0.02 &   (10)         \\
           &  105.7 &  0.31 $\:\pm$ 0.04 &   (10)         \\
           &  140.3 &  0.72 $\:\pm$ 0.07 &   (10)         \\
           &  152.6 &  0.86 $\:\pm$ 0.09 &   (10)         \\
           &  193.4 &  1.91 $\:\pm$ 0.20 &   (10)         \\
           &  205.3 &  2.19 $\:\pm$ 0.22 &   (10)         \\
           &  226.8 &  2.88 $\:\pm$ 0.29 &   (10)         \\
           &  234.1 &  3.29 $\:\pm$ 0.33 &   (10)         \\
           &  241.9 &  3.46 $\:\pm$ 0.35 &   (10)         \\
           &  250.1 &  3.94 $\:\pm$ 0.40 &   (10)         \\
           &  337.3 &  6.84 $\:\pm$ 0.68 & 2021.1.00443.S \\ 
           &  413.6 &  7.54 $\:\pm$ 0.76 &   (11)         \\
J1319+0950 &   97.0 &  0.31 $\:\pm$ 0.09 &   (12)         \\           
           &  103.4 &  0.25 $\:\pm$ 0.03 & 2018.1.01289.S \\ 
           &  233.0 &  3.89 $\:\pm$ 0.41 &   (13)         \\
           &  250.0 &  4.20 $\:\pm$ 0.77 &   (14)         \\
           &  258.0 &  5.23 $\:\pm$ 0.53 &   (14)         \\
           &  267.0 &  6.03 $\:\pm$ 0.61 &   (13)         \\
           &  306.0 &  7.39 $\:\pm$ 0.75 &   (13)         \\
           &  337.8 &  8.56 $\:\pm$ 0.86 &   (15)         \\
           &  347.0 &  9.68 $\:\pm$ 0.98 &   (13)         \\
           &  600   &       $<$6.8       &   (16)         \\
           &  857   &       $<$6.3       &   (16)         \\
           & 1200   &       $<$5.8       &   (16)         \\
P036+03    &  107.0 &  0.13 $\:\pm$ 0.02 &   (17)         \\
           &  252.0 &  2.55 $\:\pm$ 0.26 &   (6)          \\
           &  252.0 &  2.50 $\:\pm$ 0.56 &   (18)         \\
           &  332.8 &  5.00 $\:\pm$ 0.50 &   (19)         \\
           &  406.8 &  6.19 $\:\pm$ 0.88 & 2021.2.00064.S \\ 
           &  671.0 &  6.65 $\:\pm$ 1.58 & 2021.2.00151.S \\ 
\tablebreak
J2054-0005 &   92.3 &  0.08 $\:\pm$ 0.01 & 2018.1.01289.S \\ 
           &  250.0 &  2.38 $\:\pm$ 0.58 &   (20), (21)   \\
           &  263.0 &  2.98 $\:\pm$ 0.30 &   (14), (21)   \\
           &  357.2 &  6.18 $\:\pm$ 0.62 &   (22)         \\
           &  488.0 & 10.35 $\:\pm$ 1.05 &   (21)         \\
           &  600   &       $<$6.5       &   (23)         \\
           &  675.0 & 10.57 $\:\pm$ 1.46 & 2016.1.01063.S \\ 
           &  857   & 12.0  $\:\pm$ 5.0  &   (23)         \\
           & 1200   & 15.2  $\:\pm$ 5.6  &   (23)         \\
           & 1873   & 10.5  $\:\pm$ 2.3  &   (23)         \\
           & 3000   &  3.1  $\:\pm$ 1.0  &   (23)         \\
\enddata
\tablecomments{Uncertainties include an assumed 10\% absolute calibration uncertainty in addition to the statistical uncertainties. Upper limits are 1$\sigma$, and include confusion noise for SPIRE limits.}
\tablerefs{ALMA Program ID Number = This Work; (1) = \citet{li22}; (2) = \citet{yang19}; (3) = \citet{yue21}; (4) = \citet{tripodi22}; (5) = \citet{venemans18}; (6) = \citet{venemans20}; (7) = \citet{andika20}; (8) = \citet{decarli23}; (9) = \citet{pappalardo15}; (10) = \citet{pensabene21}; (11) = \citet{tripodi24}; (12) = \citet{wang11}; (13) = \citet{carniani19}; (14) = \citet{wang13}; (15) = \citet{herreracamus20}; (16) = R. Wang, priv. comm.; (17) = \citet{decarli22}; (18) = \citet{banados15}; (19) = \citet{butler23}; (20) = \citet{wang08}; (21) = \citet{hashimoto19}; (22) = \citet{salak24}; (23) = \citet{leipski14}}
\end{deluxetable}

\end{CJK*}
\end{document}